%% file: EAirQ-template.tex
\documentclass[lettersize,journal]{IEEEtran}

\IEEEoverridecommandlockouts

\pdfoutput=1

\usepackage{url}
\usepackage{multirow}
\usepackage{makecell}

\usepackage{nameref}

\usepackage{cite}
\usepackage{amsmath,amssymb,amsfonts}
\usepackage{mathtools}
\usepackage[noend]{algpseudocode}
\usepackage{algorithm}
\usepackage{algpseudocode}
\usepackage{subcaption}
\usepackage[justification=centering]{caption}

\makeatletter
\newcommand{\algmargin}{\the\ALG@thistlm}
\makeatother
\newlength{\whilewidth}
\algdef{SE}[parWHILE]{parWhile}{EndparWhile}[1]
{\parbox[t]{\dimexpr\linewidth-\algmargin}{%
		\hangindent\whilewidth\strut\algorithmicwhile\ #1\ \algorithmicdo\strut}}{\algorithmicend\ \algorithmicwhile}%
\algnewcommand{\parState}[1]{\State%
	\parbox[t]{\dimexpr\linewidth-\algmargin}{\strut #1\strut}}

\usepackage{graphicx}
\usepackage{textcomp}
\usepackage{xcolor}
\usepackage{verbatim}
\def\BibTeX{{\rm B\kern-.05em{\sc i\kern-.025em b}\kern-.08em
		T\kern-.1667em\lower.7ex\hbox{E}\kern-.125emX}}

\DeclareMathOperator{\Fb}{F_2}
\DeclareMathOperator{\Fa}{F_1}
\DeclareMathOperator{\Da}{D_1}
\DeclareMathOperator{\Db}{D_2}
\DeclareMathOperator{\Dc}{D_3}

\DeclareMathOperator{\Dis}{Dis}
\DeclareMathOperator{\Sim}{Sim}
\DeclareMathOperator{\LWD}{Lis}
\DeclareMathOperator{\Nor}{Nor}

\usepackage{float}
\newcommand{\tabincell}[2]{\begin{tabular}{@{}#1@{}}#2\end{tabular}}

\usepackage{amssymb}
\usepackage{amsmath}
\usepackage{multicol}

\usepackage{lmodern} 

\def\BibTeX{{\rm B\kern-.05em{\sc i\kern-.025em b}\kern-.08em
		T\kern-.1667em\lower.7ex\hbox{E}\kern-.125emX}}

\begin{document}

\title{Lightweight privacy-preserving truth discovery for vehicular air quality monitoring
}

\author{Rui Liu$^a$, Jianping Pan$^a$
		\thanks{$^a$Department of Computer Science, University of Victoria, Victoria, BC, V8P 5C2, Canada}
		\thanks{Rui Liu: liuuvic@uvic.ca}
}

\maketitle
\begin{abstract}

Air pollution has become a global concern for many years. Vehicular crowdsensing systems make it possible to monitor air quality at a fine granularity. To better utilize the sensory data with varying credibility, truth discovery frameworks are introduced. However, in urban cities, there is a significant difference in traffic volumes of streets or blocks, which leads to a data sparsity problem for truth discovery. Protecting the privacy of participant vehicles is also a crucial task. 
We first present a data masking-based privacy-preserving truth discovery framework, which incorporates spatial and temporal correlations to solve the sparsity problem. To further improve the truth discovery performance of the presented framework, an enhanced version is proposed with anonymous communication and data perturbation. Both frameworks are more lightweight than the existing cryptography-based methods. We also evaluate the work with simulations and fully discuss the performance and possible extensions.

\end{abstract}

\begin{IEEEkeywords}
Privacy preserving, Truth discovery, Crowdsensing, Vehicular networks
\end{IEEEkeywords}


\input{EAirQ_1_intro}

\input{EAirQ_2_pre}
\input{EAirQ_3_AirQ}

\input{EAirQ_4_EAirQ}
\input{EAirQ_5_eva}
\input{EAirQ_6_Dis}

%


\input{egbib.bbl}

\bibliographystyle{IEEEtran}    
\bibliography{egbib}
~~~\\
~~~\\







\end{document}

%% file: EAirQ_1_intro.tex
\section{Introduction}
\label{sec:intro}
Air pollution is a major health and environmental concern these years. However, performing fine-grained tasks is a challenge with the air quality monitoring stations deployed in practice.
For example, citizens would like to know the latest best route with fresh air for cycling but the stations in use are usually inadequate.
Vehicular CrowdSensing (VCS) is one possible solution to accomplish such tasks. In VCS, vehicles equipped with onboard sensors collect data from each block or street of a city. The data are uploaded to a server at periodic intervals. As a result, the server can update the air quality values at a fine granularity.

In VCS, one typical challenge is truth discovery. 
To be specific, the sensory data provided by vehicles usually vary in quality because of  different precisions of onboard sensors and possibly malicious behaviors of the drivers. 
Thus, discovering the reliability of participants which is unknown a priori from the biased or fake data is of significant importance.
The process of finding the true results of the task and the reliability of each participant is called truth discovery.

Many studies on truth discovery have been conducted in recent years~\cite{wang2014maximum, su2014generalized, li2014resolving, jin2017theseus, xu2019efficient, du2019bayesian}. These proposed approaches usually need a large amount of data to gain high accuracy.
However, in real life, only a small portion of blocks have considerably high traffic while a large number of blocks cannot provide adequate data~\cite{xu2016understanding}, which is referred to as the \emph{long tail phenomenon}. This data sparsity problem may result in inaccurate reliability discovery and truth finding. Additionally, the trajectories of vehicles, containing sensitive information such as the locations of home and companies, may be revealed in the process. Protecting the privacy is a challenge in the truth discovery of VCS.

To address the above concerns, we first propose a privacy-preserving truth discovery framework \emph{AirQ}~(\textbf{Air} \textbf{Q}uality) for vehicular crowdsensing. A truth discovery algorithm, \emph{ST}~(\textbf{S}patial and \textbf{T}emporal), is presented for AirQ to handle the data sparsity problem.
The intuitions are threefold: 1) neighbor blocks or streets are likely to have similar air quality owing to the dispersion of atmospheric pollutants. 2) The current quality value can be predicted from the historical data because the change of air quality usually takes time. 3) The historical reliability of participants can be utilized to help estimate the current reliability. In AirQ, to protect both the observation values and trajectories of vehicles, the essential data are masked before uploading. Simulation results show that AirQ works well in fine-grained air quality monitoring, especially when the provided data are insufficient. 

To further improve the performance of AirQ, we propose an enhanced version, named as \emph{EAirQ}~(\textbf{E}nhanced \textbf{AirQ}). The ST algorithm used in AirQ is simplified for truth discovery with sufficient data. However, the masking technique cannot be well adopted in EAirQ to preserve the privacy. 
To tackle the challenge, 1) a new framework architecture is developed, where an anonymous authentication scheme is adopted; 2) a two-layer perturbation scheme is presented inspired by the idea of randomized response and local differential privacy. Simulation results show that EAirQ has a better performance of truth discovery compared to AirQ while also maintaining the privacy-preserving property.

The main work and contributions of the paper are as follows:
\begin{itemize}
	\item We present an optimization-based truth discovery algorithm, ST. Spatial and temporal correlations are combined to solve the data sparsity problem, which makes the algorithm suitable for fine-grained tasks.
	\item We present a privacy-preserving framework, AirQ, based on the technique of masking. We circumvent the limitations of masking and protect the observation values and the trajectories of vehicles in the process of truth discovery. 
	\item We further develop EAirQ from AirQ with the techniques of anonymous authentication and perturbation. Two different truth discovery algorithms are combined to reduce the negative impact of data reuse when there are sufficient data.
	\item Different from the existing privacy-preserving methods based on cryptography, both AirQ and EAirQ are lightweight from the perspective of computation and communication costs on vehicles. Thus, they are suitable for vehicular networks.
\end{itemize}

The paper is structured as follows. In Section~\ref{sec:pre}, we show the related work, introduce the cryptography techniques and define the problem. The details of the proposed ST algorithm and AirQ framework are provided in Section~\ref{sec:fra_AirQ}. The enhanced framework EAirQ is introduced in Section~\ref{sec:fra_EAirQ}. We conduct a series of simulations and analyze the privacy of our proposed frameworks in Section~\ref{sec:eva}. In Section~\ref{sec:dis}, the possible scenarios, the extensions and the remaining issues are discussed. We conclude the paper in Section~\ref{sec:con}.

%% file: EAirQ_2_pre.tex
\section{Preliminaries}
\label{sec:pre}
   
\subsection{Related work}
Truth discovery has become a hot topic these years. To solve the sparsity problem, Zhang~{\em et al.}~\cite{zhang2016robust} presented a robust truth discovery scheme which quantifies the attitude that human expressed and incorporates the historical contributions. Yang~{\em et al.}~\cite{yang2019using} incorporated the information about the social network in a truth discovery framework and developed Laplace variational inference methods to estimate participants' reliabilities. However, these schemes focus on social network. Because of the different demands and challenges, they cannot well adapt to fine-grained air quality monitoring.

Cryptography has been adopted to address the privacy issues in crowdsensing~\cite{xu2017achieving,blasco2015three,wang2018reliable}. Miao~{\em et al.}~\cite{miao2015cloud} proposed a mechanism called PPTD based on the threshold Paillier cryptosystem. The proposed scheme can preserve the privacy of weights and observation values in truth discovery. However, considerable amounts of cryptography-based calculations have to be conducted by participants, which is a common limitation of cryptosystem-based schemes.

\subsection{Cryptography tools}
In this section, we briefly introduce the related cryptography tools and techniques.
\subsubsection{Data masking}
\label{cry_datamasking}
 Data masking allows a server to aggregate data from client parties in a secure way. In an additive masking algorithm, all sensitive inputs are masked by adding random values called \emph{masks}. The randomness should be canceled once the masked inputs are aggregated. Thus, the server only learns the sum of the clients' inputs.
 
 In this paper, we adopt a masking algorithm with one-time pads~\cite{bonawitz2017practical}. In this algorithm, suppose the set of client parties is $U$.
 An input from client $a\in U$ is denoted as $x_a$. Each pair of clients $(a,b)$ that satisfy $a<b$ agrees on a mask $\alpha_{a,b}$. Note that we use $a<b$ to represent that the index of client $a$ is smaller than that of $b$ for simplicity. Then the masked value of $x_a$ can be represented as:
\begin{equation}
\label{equ_masking}
y_a = x_a +\sum_{b\in U: a<b}\alpha_{a,b} - \sum_{b\in U: a>b}\alpha_{b,a}
\end{equation}

After collecting all the masked values, the server can compute the sum of $\{x_a\mid a\in U\}$ as follows:
\begin{equation}
\label{equ_unmasking}
\begin{split}
\sum_{a\in U} y_a =& \sum_{a\in U}\Bigl(x_a +\sum_{b\in U: a<b}\alpha_{a,b} - \sum_{b\in U: a>b}\alpha_{b,a}\Bigr)\\
=&\sum_{a\in U} x_a
\end{split}
\end{equation}

In the algorithm of~\cite{bonawitz2017practical}, each pair of clients $(a,b)$ should exchange secrets to reach an agreement on the mask $\alpha_{a,b}$, which brings high communication cost. Besides, drops of clients after masking will result in failure of summation. However, these are not problems in our work, which will be described in details in Section~\ref{sec:fra_AirQ}.

\subsubsection{Anonymous authentication}
\label{subsec:pre_anoy}
A digital signature-based anonymous authentication scheme provides not only the integrity and authenticity of a message, but also the anonymity of the signer. 
It can be achieved by using pseudo-IDs. To be specific, a message sender (or signer) uses a pseudo-ID in the signing process issued by a trusted third party as a replacement of the real ID. The message receiver can verify the message sent but cannot trace the real identity of the sender.

Moni~{\em et al.}~\cite{moni2020scalable} presented a distributed, scalable and low-overhead authentication scheme for VANETs~(Vehicular Ad hoc Networks). The scheme has two layers: the upper layer of the Trusted Authority (TA) and Regional Trusted Authorities (RTAs), and the lower layer of vehicles and RSUs.
In this work, vehicles are issued with pseudo-IDs by RTAs. Messages are then signed with the pseudo-IDs by the RSA algorithm. Only TA and RTAs have the ability to reveal the real identities of vehicles. 
In our work, the scheme is adopted to achieve the anonymous communication between a vehicle and an RSU. We omit the description of the parameter distribution, the communication establishment and the signature construction in this paper for brevity. For more details, 
we refer readers to \cite{moni2020scalable}.

\subsubsection{Randomized response and local differential privacy}
Randomized response is a survey technique that allows surveyees to respond to sensitive questions while maintaining the confidentiality. A randomization device (e.g., a coin flip) is used by surveyees to decide if an answer should be given truthfully. The interviewer can get a reliable statistic result from the biased answers. 

Local Differential Privacy (LDP) is a model to protect individuals' privacy in statistical computations. Different from differential privacy, there is no trusted central server (i.e., a data collector) in LDP because participants perturb the raw data locally. 

Although we do not adopt any specific randomized response or LDP algorithms in our work, the ideas extracted from the two techniques are used to develop a perturbation mechanism. Details can be found in Section~\ref{subsec:perturbation}.

\subsection{Problem definition}
\label{sebsec:prb_def}
In crowdsensing systems, there are usually two types of parties:  \emph{sources} and a \emph{server}. Sources are the participants who conduct sensing tasks and then upload the sensory data to a server for further processing and analysis. The sensory readings for a specific sensing task are called \emph{observation values}. The actual true value of a task is denoted by \emph{ground truth}.
In truth discovery algorithms, \emph{weight} represents the reliability of a source. \emph{Truth} denotes the estimated ground truth of a task based on the collected sensory data and weights.

We formally define the problem targeted as follows:

We divide the urban area to $m$ disjoint \emph{grids} $G=\{g_1, g_2, \dots, g_m\}$, typically streets or blocks in practice. A \emph{sensing cycle} is a static time slot (e.g., 15 minutes). In each sensing cycle, the sensory data are uploaded once. Then the truths and weights can be updated based on the data. 
The specific crowdsensing task is to get an estimated air quality value for each grid in each sensing cycle.

There are $n$ sources, i.e., vehicles, registered in the system denoted by $S=\{s_1, s_2, \dots, s_n\}$. Note that we use the terms ``source" and ``vehicle" interchangeably in the following sections. A source $s$ provides a \emph{report}, containing a certain number (denoted by $c$) of observation values, $V_s=\{v_{s,1}, v_{s,2},\dots, v_{s,c}\}$, in each sensing cycle. The $j$-th observation value is denoted as $v_{s,j}$ where $j\in\{1,2,\dots,c\}$. 
The grid where $v_{s,j}$ is generated is denoted as $g^{s,j}$. We assume that each source provides at most one observation value for each grid.
The weight of $s$ at sensing cycle $t$ is denoted as $w_{s,t}$, which combines the temporary weight $w_{s,t}^{\prime}$ and the historical weights $W_s = \{w_{s,1}, w_{s,2}, \dots, w_{s,t-1}\}$.
Similarly, the estimated ground truth of $g$ at sensing cycle $t$, denoted as $v_{g,t}^*$, combines the temporary truth $v_{g,t}^{*\prime}$ and the historical truths $T_g=\{v_{g,1}^*, v_{g,2}^*, \dots, v_{g,t-1}^*\}$.
Note that for simplicity, we omit some subscripts. For example, we use $s$ to denote $s_i$ where $i\in\{1,2,\dots,n\}$.
Our goal is to let the server calculate $v_{g,t}^*$ for each $g$ and $w_{s,t}$ for each $s$ in each sensing cycle $t$. 


We assume that all parties are semi-honest. To be specific, all the parties follow the protocol of the proposed frameworks but may try to infer the sensitive information of other parties from the reports. 
The privacy-preserving goal in this paper is that any observation value $v_{s,j}$ and the trajectory of vehicle $s$ should not be revealed from the reports to any party except $s$ itself. One point worth mentioning is that EAirQ introduces a new entity, a trusted manager. It manages all vehicles and is considered to be honest and trustworthy. More introduction is given in Section~\ref{subsec: EAirQ}.

In addition, we assume the communications among all parties are reliable. All packets can be sent and received successfully in the network.
The communication performance and the Quality of Service (QoS), such as the packet loss rate, are addressed in the lower layers. Thus, we can focus on the truth discovery performance and the security and privacy-preservation goals in the application layer.
%
%

%% file: EAirQ_3_AirQ.tex
\section{The AirQ framework}
\label{sec:fra_AirQ}
In this section, we first introduce the proposed truth discovery algorithm, ST, which tackles the long tail phenomenon of vehicular networks. Then the privacy-preserving AirQ framework is described in details.

\subsection{Truth discovery algorithm}
\label{subsec:td_alg}
To address the data sparsity problem, we take the spatial correlation of grids and the temporal correlations of weights and truths into consideration. We first introduce the details of the correlations and then propose the optimization problem.

\subsubsection{Spatial correlation}
When calculating the estimated ground truth $v_{g,t}^*$, not only the observation values provided for $g$ but also the values for other grids are used. The correlation between two grids is represented by a parameter $\theta_{s,j,g}$, i.e., how much we can rely on $v_{s,j}$ for $v_{g,t}^*$. The intuition is that the nearer the two grids are, the more likely they have similar air quality. Thus, $\theta_{s,j,g}$ is calculated by the \emph{logical distance} $\Dis(g^{s,j}, g)$ of the two grids $g^{s,j}$ and $g$. We adopt the \emph{Gaussian kernel} for $\Dis(g^{s,j}, g)$ as follows~\cite{meng2015truth,li2018holmes}:
\begin{equation}
\label{equ_gaussian}
\Dis(g^{s,j}, g) =
\left\{
\begin{array}{lr}
\exp(-\frac{\Da(g^{s,j}, g)^2}{2\omega^2}), &\text{if}\; \Da(g^{s,j}, g)< u \\
0, & \text{otherwise}
\end{array} \right.
\end{equation}
where $\omega$ is the width parameter of the kernel and $u$ is a threshold we set.
$\Da(g^{s,j}, g)$ is the \emph{geographical distance} between $g^{s,j}$ and $g$~\cite{liu2020airq}.


\subsubsection{Temporal correlation}
A source who performed bad, i.e., had low weights, in the past, is likely to be unreliable in the current sensing cycle. In other words, the historical weights of a source can be used to predict the latest weight. Based on this intuition, we define $w_{s,t}$ as
\begin{equation} 
\label{equ_tempweight}
\begin{split}	 
w_{s,t} &= \Fa(w_{s,t}^{\prime}, W_s)\\
& =  
\left\{
\begin{array}{lr}
\frac{\sum_{i=1}^{t-1}{k_i w_{s,i}}+k_t w_{s,t}^{\prime}}{\sum_{i=1}^{t} k_i}, &\text{if}\; W_s\neq \emptyset \\
w_{s,t}^{\prime}, &\text{otherwise}
\end{array}\right.
\end{split}
\end{equation}
where the function $\Fa$ combines the historical weights $W_s$ and the temporary weight $w_{s,t}^{\prime}$ of source $s$ with the \emph{inverse distance weighting} method. $k_i$ is defined as:
\begin{equation} 
k_i=\frac{1}{\Db(w_{s,t}, w_{s,i})^{\rho_w}}
\end{equation}

$\rho_w$ is a positive real number called the power parameter. It controls the degree of dependence on historical weights. $\Db(w_{s,t}, w_{s,i})$ is the \emph{temporal distance}  between two data points, i.e., $w_{s,t}$ and $w_{s,i}$~\cite{liu2020airq}. In other words, it represents the time interval between the past and current sensing cycles. 

The parameter $k_i$ results in a negative correlation between the temporal distances and the significance of past weights in weight updating.

Similarly, the ground truths of grids are not only related to the observation values but also the past records because the easing of air pollution takes time. Based on this intuition, we define $v_{g,t}^*$ as:
\begin{equation} 
\label{equ_temptruth}
\begin{split}	 
v_{g,t}^{*} &= \Fb(v_{g,t}^{*\prime}, T_g)\\
& = 
\left\{
\begin{array}{lr}
\frac{\sum_{i=1}^{t-1}{k_i^{\prime} v_{g,i}^{*}}+k_t^{\prime} v_{g,t}^{*\prime}}{\sum_{i=1}^{t} k_i}, &\text{if}\; T_g\neq \emptyset \\
v_{g,t}^{*\prime}, &\text{otherwise}
\end{array}\right.
\end{split}
\end{equation}
where
\begin{equation} 
\label{equ_truthcorr}
k_i^{\prime}=\frac{1}{\Db(v_{g,t}, v_{g,i})^{\rho_t}}
\end{equation}

The power parameter $\rho_t$ controls the degree of dependence on historical truths. With $k_i^{\prime}$, the newer historical truths are more significant to the estimation of the current truth for each grid. For simplicity, we transform Equations~\eqref{equ_tempweight} and~\eqref{equ_temptruth} to:
\begin{equation}
\label{equ_delta12}
v_{g,t}^{*}=\delta_{1,g}+\delta_{2,g} v_{g,t}^{*\prime}
\end{equation}
\begin{equation}
\label{equ_delta34}
w_{s,t}=\delta_{3,s}+\delta_{4,s} w_{s,t}^\prime
\end{equation}

\subsubsection{Optimization problem}
\label{subsubsec:opti}
Taking full advantage of the above correlations, we formulate the optimization problem for truth discovery as follows:
\begin{equation}
\label{equ_opti}
\begin{split}
&\min_{\{w_{s,t}^\prime\},\{v_{g,t}^{*\prime}\}}\sum_{s\in S}\sum_{g\in G}\sum_{j\in\{1,2,\dots, c\}}
\Fa(w_{s,t}^{\prime},W_s) \theta_{s,j,g}\\
&\Dc(v_{s,j},\Fb(v_{g,t}^{*\prime}, T_g)),\\
&\text{s.t.} \sum_{s\in S}\exp(-\Fa(w_{s,t}^\prime, W_s))=1
\end{split}
\end{equation}
where $\theta_{s,j,g}$ controls the reuse of sensory data. Functions $\Fa$ and $\Fb$ combine the historical records with the temporary results.
$\Dc(v_{s,j},\Fb(v_{g,t}^{*\prime}, T_g))$ is the deviation between an observation value $v_{s,j}$ and the estimated ground truth $\Fb(v_{g,t}^{*\prime}, T_g)$ (or $v_{g,t}^{*}$). We use \emph{truth distance} to represent the deviation. The squared L2-norm is adopted to calculate it. The constraint regularizes the value of $w_{s,t}$ by constraining the sum of $\exp(-w_{s,t})$~\cite{li2014resolving}. 

Solving the above convex optimization problem by KKT (Karush–Kuhn–Tucker) conditions, we have:
\begin{equation}
\label{equ_truth_updating}
v_{g,t}^{*\prime}=\frac{\sum_{s\in S}\sum_{j\in\{1,2,\dots, c\}}\Fa(w_{s,t}^{\prime},W_s)\theta_{s,j,g}(v_{s,j}-\delta_{1,g})}{\delta_{2,g} \sum_{s\in S}\sum_{j\in\{1,2,\dots, c\}} \Fa(w_{s,t}^{\prime},W_s)\theta_{s,j,g}}
\end{equation}
\begin{equation}
\label{equ_weight_updating}
\begin{split}
&w_{s,t}^\prime=\\
&\frac{1}{\delta_{4,s}}\bigl( \log 
{\left(
	\frac{ \splitfrac{\sum_{s\in S}\sum_{g\in G}\sum_{j\in\{1,2,\dots, c\}}\theta_{s,j,g} }
		{\| v_{s,j}-\Fb(v_{g,t}^{*\prime},T_g)\|^2}}
	{ \splitfrac {\sum_{g\in G}\sum_{j\in\{1,2,\dots, c\}}\theta_{s,j,g}}
		{\| v_{s,j}-\Fb(v_{g,t}^{*\prime},T_g)\|^2}} 
	\right)}
-\delta_{3,s} \bigr)
\end{split}
\end{equation}

Based on the solution above, we summarize the ST algorithm in Algorithm~\ref{alg_td}. Readers can refer to the previous work~\cite{liu2020airq} for more details.

\begin{algorithm}[hbt!]
	\caption{Truth discovery algorithm: ST}
	\label{alg_td}
	\textbf{Input}: Observation values from $n$ sources: $\{V_s\mid  s\in S\}$, historical truths of $m$ grids: $\{T_g\mid g\in G\}$, historical weights from $n$ sources: $\{W_s\mid  s\in S\}$, and parameters: $\{\theta_{s,j,g}\mid  s\in S, j\in\{1,2,\dots,c\}, g\in G\}$\\ 
	\textbf{Output}: Estimated ground truths for $m$ grids: $\{v_{g,t}^*\}$, weights for $n$ sources: $\{w_{s,t}\}$, and updated records: $\{T_g\mid g\in G\}$ and $\{W_s\mid  s\in S\}$
	\begin{algorithmic}[1]
		\State Initialize $v_{g,t}^{*\prime}$ for each grid $g$ to the average of all the observation values provided for the grid;
		\State Initialize $w_{s,t}^\prime$ for each source $s$ to $\frac{1}{n}$;
		\State Calculate $\delta_{1,g}$ and $\delta_{2,g}$ based on $W_s$ for each source $s$ (i.e., Equation~\eqref{equ_tempweight} and \eqref{equ_delta12});
		\State Calculate $\delta_{3,s}$ and $\delta_{4,s}$ based on $T_g$ for each grid $g$ (i.e., Equation~~\eqref{equ_temptruth} and \eqref{equ_delta34});
		\Repeat 
		\For{each source $s$} 
		\parState{Update $w_{s,t}^\prime$ based on $\delta_{3,s}$, $\delta_{4,s}$, $\{\theta_{s,j,g}\mid  s\in S, j\in\{1,2,\dots,c\}, g\in G\}$, $\{V_s\mid  s\in S\}$ and $\{v_{g,t}^{*\prime}\}$~(i.e., Equation~\eqref{equ_weight_updating});}
		\EndFor
		\For{each grid $g$} 
		\parState{Update $v_{g,t}^{*\prime}$ based on $\delta_{1,g}$, $\delta_{2,g}$, $\{\theta_{s,j,g}\mid  s\in S, j\in\{1,2,\dots,c\}\}$, $\{V_s\mid  s\in S\}$ and $\{w_{s,t}^{\prime}\}$~(i.e., Equation~\eqref{equ_truth_updating});}
		\EndFor
		\Until {the convergence criterion is satisfied};
		\parState {Update $v_{g,t}^*$ for each $g$ based on $v_{g,t}^{*\prime}$, $\delta_{1,g}$ and $\delta_{2,g}$~(i.e., Equation~\eqref{equ_delta12});}
		\parState {Update $w_{s,t}$ for each $s$ based on $w_{s,t}^{\prime}$, $\delta_{3,s}$ and $\delta_{4,s}$~(i.e., Equation~\eqref{equ_delta34});}
		\parState {Append $v_{g,t}^*$ to $T_g$ for each $g$;}
		\parState {Append $w_{s,t}$ to $W_s$ for each $s$;}
		\Return $\{v_{g,t}^*\}$, $\{w_{s,t}\}$, $\{T_g\mid g\in G\}$ and $\{W_s\mid  s\in S\}$
	\end{algorithmic}
\end{algorithm}

\subsection{AirQ framework}
To address the privacy issues, we develop the ST algorithm into the AirQ framework with the technique of masking. The working principle of AirQ is introduced in this section.

\subsubsection{Data generation}
\label{subsub:datagene}
Vehicles generate the observation values for the grids they pass by. Meanwhile, a vehicle $s$ should ask for $\{\theta_{s,j,g}\mid  g\in G\}$ from the nearest RSU for each $v_{s,j}$. 
Recall that $\theta_{s,j,g}$ represents the spatial correlation between $g^{s,j}$ and $g$, which is a constant. Thus, recording all the $\{\theta_{s,j,g}\mid  g\in G\}$ by the nearest RSU of grid $g^{s,j}$ reduces the storage burden on vehicles. Besides, $\theta_{s,j,g}$ equals 0 when the two grids are far from each other and have an insignificant spatial correlation. Therefore, RSUs only need to send the none-zero values to lower the communication cost.

 \subsubsection{Data masking}
\label{subsec_datamasking}
Before uploading reports in every sensing cycle,
each vehicle $s$ masks three types of values $\{\theta_{s,j,g}\cdot v_{s,j}\mid  j\in\{1,2,\dots,c\}\}$, $\{\theta_{s,j,g}\cdot v_{s,j}^2\mid  j\in\{1,2,\dots,c\}\}$, and $\{\theta_{s,j,g}\mid  j\in\{1,2,\dots,c\}\}$ for each grid $g$ by the masking algorithm with one-time pads~\cite{bonawitz2017practical}. In this paper, a significant difference is that each vehicle $s$ chooses masks for pairs of specific values it maintains. Thus, there is no need for secret exchange protocols when masking and drops of clients will not impede the truth discovery process. The difference guarantees the low computation and communication costs on the vehicle-side and the availability of the framework. 

We use $\{\theta_{s,j,g}\cdot v_{s,j}\mid  j\in\{1,2,\dots,c\}\}$ as an example to describe the process in details. 
For each pair of $(\theta_{s,j,g}\cdot v_{s,j}, \theta_{s,j^\prime,g}\cdot v_{s,j^\prime})$ that satisfies $j<j^\prime$, $s$ generates a random value $\alpha_{j,j^\prime}^{s,g}$ by a Pseudo-Random Number Generator~(PRNG). 
We use $\beta_{1}^{s,j,g}$ to denote the masked $\theta_{s,j,g}\cdot v_{s,j}$. It can be calculated based on Equation~\eqref{equ_masking} as follows:
\begin{equation}
\label{equ_maskingour01}
\begin{split}
\beta_{1}^{s,j,g} = & \theta_{s,j,g}\cdot v_{s,j} +\sum_{j^\prime \in \{1,2,\dots,c\}: j<j^\prime}\alpha_{j,j^\prime}^{s,g} \\
& - \sum_{j^\prime \in \{1,2,\dots,c\}: j>j^\prime}\alpha_{j^\prime,j}^{s,g}
\end{split}
\end{equation}

Based on Equation~\eqref{equ_unmasking}, $ \chi_1^{s,g}$, the sum of $\{\theta_{s,j,g}\cdot v_{s,j}\mid  j\in\{1,2,\dots,c\}\}$ can be calculated as:
\begin{equation}
\label{equ_sum01}
\begin{split}
\chi_1^{s,g} &= \sum_{j\in\{1,2,\dots,c\}} \theta_{s,j,g}\cdot v_{s,j} \\
&= \sum_{j\in\{1,2,\dots,c\}} \beta_{1}^{s,j,g}
\end{split}
\end{equation}

Similarly, we denote the masked $\theta_{s,j,g}\cdot v_{s,j}^2$ as $\beta_{2}^{s,j,g}$ and the masked $\theta_{s,j,g}$ as $\beta_{3}^{s,j,g}$. With the masking, the server can calculate the corresponding sums without learning the unmasked data. More details are given in the preliminary version~\cite{liu2020airq}. 

\subsubsection{Data uploading}
\label{subsubsec:datauploading}
In each sensing cycle, a vehicle $s$ only uploads data once to the server. The uploaded report contains $\{\beta_{1}^{s,j,g}\mid j\in\{1,2,\dots,c\}\}$, $\{\beta_{2}^{s,j,g}\mid j\in\{1,2,\dots,c\}\}$ and $\{\beta_{3}^{s,j,g}\mid j\in\{1,2,\dots,c\}\}$ for each grid $g\in G$. To reduce the communication cost, $s$ does not transfer the report to the server directly but to the nearest RSU at that time. The server then collects all the reports from all the RSUs.

\subsubsection{Data handling}
\label{subsec:datahandling}

Based on Equation~\eqref{equ_unmasking},  Equations~\eqref{equ_truth_updating} and~\eqref{equ_weight_updating} can be transformed as follows:
\begin{equation}
\label{equ_truth_updating_transfer}
\begin{split}
v_{g,t}^{*\prime} &= \frac{\sum_{s\in S}\sum_{j\in\{1,2,\dots, c\}}\Fa(w_{s,t}^{\prime},W_s)\theta_{s,j,g}(v_{s,j}-\delta_{1,g})}{\delta_{2,g} \sum_{s\in S}\sum_{j\in\{1,2,\dots, c\}} \Fa(w_{s,t}^{\prime},W_s)\theta_{s,j,g}}\\
&= \frac{\sum_{s\in S}\Fa(w_{s,t}^{\prime},W_s)\bigl( \chi_1^{s,g}- \delta_{1,g}\chi_3^{s,g}\bigr)}
{\delta_{2,g} \sum_{s\in S} \Fa(w_{s,t}^{\prime},W_s) \chi_3^{s,g}}
\end{split}
\end{equation}

 \begin{equation}
\label{equ_weight_updating_transfer}
\begin{split}
w_{s,t}^\prime=&
\frac{1}{\delta_{4,s}}\bigl( \log 
{\left(
	\frac{ \splitfrac{\sum_{s\in S}\sum_{g\in G}\sum_{j\in\{1,2,\dots, c\}}\theta_{s,j,g} }
		{\| v_{s,j}-\Fb(v_{g,t}^{*\prime},T_g)\|^2}}
	{ \splitfrac {\sum_{g\in G}\sum_{j\in\{1,2,\dots, c\}}\theta_{s,j,g}}
		{\| v_{s,j}-\Fb(v_{g,t}^{*\prime},T_g)\|^2}} 
	\right)}\\
&-\delta_{3,s}\bigr)\\
=&\frac{1}{\delta_{4,s}}\bigl( \log 
{\left(
	\frac{\splitfrac{\sum_{s\in S}\sum_{g\in G}
			\chi_2^{s,g}
			- 2 \Fb(v_{g,t}^{*\prime},T_g)}
		{ \cdot\chi_1^{s,g} + \Fb(v_{g,t}^{*\prime},T_g)^2 \chi_3^{s,g}}}
	{ \splitfrac{ \sum_{g\in G}
			\chi_2^{s,g} - 2 \Fb(v_{g,t}^{*\prime},T_g)}
		{\cdot\chi_1^{s,g} + \Fb(v_{g,t}^{*\prime},T_g)^2\chi_3^{s,g}}
	} 
	\right)}\\
&-\delta_{3,s}\bigr)
\end{split}
\end{equation}

Therefore, the server can update $\{v_{g,t}^*\}$ and $\{w_{s,t}\}$ only with the collected $\{\beta_{1}^{s,j,g}\mid j\in\{1,2,\dots,c\}, g \in G\}$, $\{\beta_{2}^{s,j,g}\mid j\in\{1,2,\dots,c\}, g \in G\}$ and $\{\beta_{3}^{s,j,g}\mid j\in\{1,2,\dots,c\}, g \in G\}$ from each vehicle $s$. In other words, both the observation values $\{V_s\}$ and $\{\theta_{s,j,g}\}$ which maintain the information of the vehicle trajectories, are not revealed to the server and RSUs. More details are discussed in Section~\ref{subsec:privacy}. 
Recall that ST requires an average of all observation values provided for a grid as the initialization ground truth~(i.e., Step 1 in Algorithm~\ref{alg_td}), which obviously cannot be calculated with the uploaded data. The historical ground truth or a random value can be used as a substitution. 

After updating $\{v_{g,t}^*\}$ and $\{w_{s,t}\}$ for the current sensing cycle, the server appends records as described in Algorithm~\ref{alg_td}.

%% file: EAirQ_4_EAirQ.tex
\section{The EAirQ framework}
\label{sec:fra_EAirQ}

AirQ is proposed to solve the data sparsity problem while preserving the privacy. 
However, when there are sufficient data, the data reuse may bring a negative impact to the truth discovery performance. This observation is discussed in Section~\ref{subsec:tdperformence_AirQ}.
In this section, we present an enhanced version of AirQ, EAirQ.

\subsection{Simplified truth discovery algorithm}
\label{subsec: SST}
\begin{algorithm}[hbt!]
	\caption{Truth discovery algorithm: SST}
	\label{alg_td2}
	\textbf{Input}: Pairs of observation values and corresponding grids from $n$ sources: $\{(v_{s,j},g^{s,j})\mid s\in S, j\in \{1,2,\dots, c\}\}$, and historical weights of $n$ sources: $\{W_s\mid  s\in S\}$\\ 
	\textbf{Output}: Estimated ground truths for $m$ grids: $\{v_{g,t}^*\}$, weights for $n$ sources: $\{w_{s,t}\}$, and updated records: $\{T_g\mid g\in G\}$ and $\{W_s\mid  s\in S\}$
	\begin{algorithmic}[1]
		\State Initialize $v_{g,t}^{*}$ for each grid $g$ to the average of all the observation values provided for the grid;
		\State Initialize $w_{s,t}^\prime$ for each source $s$ to $\frac{1}{n}$;
		\State Calculate $\delta_{3,s}$ and $\delta_{4,s}$ based on $T_g$ for each grid $g$ (i.e., Equation~~\eqref{equ_temptruth} and \eqref{equ_delta34});
		\Repeat 
		\For{each source $s$} 
		\parState{Update $w_{s,t}^\prime$ based on $\delta_{3,s}$, $\delta_{4,s}$, and $\{(v_{s,j},g^{s,j})\mid j\in \{1,2,\dots, c\}\}$~(i.e., Equation~\eqref{equ_weight_updating2});}
		\EndFor
		\For{each grid $g$} 
		\parState{Update $v_{g,t}^{*}$ based on $\{(v_{s,j},g^{s,j})\mid s\in S, j\in \{1,2,\dots, c\}\land g^{s,j}=g\}$ and $\{w_{s,t}^{\prime}\}$~(i.e., Equation~\eqref{equ_truth_updating2});}
		\EndFor
		\Until {the convergence criterion is satisfied};
		\parState {Update $w_{s,t}$ for each $s$ based on $w_{s,t}^{\prime}$, $\delta_{3,s}$ and $\delta_{4,s}$~(i.e., Equation~\eqref{equ_delta34});}
		\parState {Append $v_{g,t}^*$ to $T_g$ for each $g$;}
		\parState {Append $w_{s,t}$ to $W_s$ for each $s$;}
		\Return $\{v_{g,t}^*\}$, $\{w_{s,t}\}$, $\{T_g\mid g\in G\}$ and $\{W_s\mid  s\in S\}$
	\end{algorithmic}
\end{algorithm}

We simplify the ST truth discovery algorithm to  \emph{SST}~(\textbf{S}implified \textbf{ST}) as a substitution of ST when there are sufficient reports. 
The corresponding optimization problem is defined as:
\begin{equation}
\label{equ_opti2}
\begin{split}
&\min_{\{w_{s,t}^\prime\},\{v_{g,t}^*\}}\sum_{s\in S}\sum_{g\in G}\sum_{j\in\{1,2,\dots, c\}\land g^{s,j}=g}
\Fa(w_{s,t}^{\prime},W_s) \\&  \Dc(v_{s,j},v_{g,t}^*),\\
&\text{s.t.} \sum_{s\in S}\exp(-\Fa(w_{s,t}^\prime, W_s))=1
\end{split}
\end{equation} 

In this problem, we only involve the reliability of a report (i.e., the weight of the data provider $w_s$) and the temporal correlation of weights (i.e., the historical weights $W_s$). Reports are not reused among grids. The historical truths are not considered as well.
The intuition is in two aspects: 1) when there are sufficient reports, the average value can reflect the ground truth accurately. 2) AirQ may cause a deviation from the average because of the incorporation of the correlations. 

Solving the above convex optimization problem by KKT conditions, we have:
\begin{equation}
\label{equ_truth_updating2}
v_{g,t}^{*}=\frac{\sum_{s\in S}\sum_{j\in\{1,2,\dots, c\}\land g^{s,j}=g}\Fa(w_{s,t}^{\prime},W_s)v_{s,j}}{\sum_{s\in S} \Fa(w_{s,t}^{\prime},W_s)}
\end{equation}

\begin{equation}
\label{equ_weight_updating2}
\begin{split}
&w_{s,t}^\prime=\\
&\frac{1}{\delta_{4,s}}\bigl( \log 
{\left(
	\frac{ \splitfrac{\sum_{s\in S}\sum_{g\in G}\sum_{j\in\{1,2,\dots, c\}\land g^{s,j}=g} }
		{\| v_{s,j}-v_{g,t}^*\|^2}}
	{ \splitfrac {\sum_{g\in G}\sum_{j\in\{1,2,\dots, c\}\land g^{s,j}=g}}
		{\| v_{s,j}-v_{g,t}^*\|^2}} 
	\right)} \\
&
-\delta_{3,s} \bigr)
\end{split}
\end{equation}

The SST algorithm is described in Algorithm~\ref{alg_td2}. In each sensing cycle $t$, $\{w_{s,t}^{\prime}\}$ and $\{v_{g,t}^*\}$ are updated by Equations~\eqref{equ_truth_updating2} and~\eqref{equ_weight_updating2} iteratively until the convergence criterion is satisfied~(i.e., Step 5--9). Final values $\{w_{s,t}\}$ and $\{v_{g,t}^{*}\}$ are appended to $W_s$ and $T_g$, respectively~(i.e., Step 10--12). 
Note that although the historical truths $\{T_g\mid g\in G\}$ are not used in SST, they are necessary to be recorded for the EAirQ framework. More details are given in Section~\ref{subsec: EAirQ}.

Different from ST, the pairs of observation values and corresponding grids $\{(v_{s,j},g^{s,j})\mid j\in \{1,2,\dots, c\}\}$ are uploaded by each $s$. It may reveal both the real observation values and the trajectories.
The masking technique adopted for ST is not suitable for SST because the inputs of the algorithms are different. It is a challenge to preserve the privacy in SST while keeping the framework as lightweight as possible.

\subsection{Perturbation mechanism}
\label{subsec:perturbation}
To overcome the new privacy challenge discussed in Section~\ref{subsec: SST}, we present a new mechanism for SST, inspired by the idea of randomized response and LDP. The mechanism adds a two-layer perturbation to the raw data as follows:

\textbf{Grid perturbation}: similar to the idea of randomized response, vehicles do not always truthfully provide the trajectories. In other words, each vehicle perturbs the records of grids it passed by as follows: 1) for each observation value $v_{s,j}$ provided by $s$ in sensing cycle $t$, $s$ removes it from the list $V_s$ with a probability $p_1$ (e.g., 0.2), as defined in Equation~\eqref{equ_rand_respon}. 
2) For each grid $g$ that satisfies $\{g^{s,j}\neq g \mid j\in \{1,\dots,c\}\}$, $s$ adds $v_{s,c+1}$ to $V_s$ with the probability $p_2$. $v_{s,c+1}$ is calculated by Equation~\eqref{equ_rand_respon2}. 

\begin{equation}
\label{equ_rand_respon}
v_{s,j} = 
\left\{
\begin{array}{lr}
v_{s,j}, &\text{with probability}\; 1-p_1  \\
\emptyset, &\text{with probability}\; p_1
\end{array}\right.
\end{equation}

\begin{equation}
\label{equ_rand_respon2}
v_{s,c+1} =  v^*_{g,t-1}+\psi_1 
\end{equation}

$c$ is the current number of reports in $V_s$. $\psi_1$ is a Laplace noise generated from a Laplace distribution $\mathcal{L}(0, \lambda_1)$ where 0 is the location parameter and $\lambda_1$ is the scale parameter. Note that we suggest setting $p_2$ with a small value (such as 0.05) to reduce the impact on accuracy. In the following parts, we use the term, \emph{imitated reports}, to denote the reports generated and added in the grid perturbation process.

\textbf{Value perturbation}: similar to the idea of LDP, each vehicle locally perturbs the observation values it provides. To be specific, for each $v_{s,j} \in V_s$, $s$ adds a Laplace noise as follows:
\begin{equation}
\label{equ_noise}
\hat{v_{s,j}} =  v_{s,j}+\psi_2 
\end{equation}
where $\psi_2 \sim \mathcal{L}(0, \lambda_2)$ and $\lambda_2$ is the scale parameter. 

In our work, the perturbation mechanism may involve bias to the truth discovery results. However, a moderate sacrifice of precision is acceptable when there are sufficient data. 
Besides, the privacy and precision can be balanced by adjusting the parameters based on different user demands and scenarios.
More discussion and analysis are given in Section~\ref{sec:eva}, where we show that the goal of EAirQ (i.e., preserving the privacy efficiently while gaining a better truth discovery performance than AirQ) is achieved. 

To provide further protection of the privacy, besides the perturbation scheme, we adopt the anonymous communication between vehicles and RSUs. More details can be found in Section~\ref{subsec: EAirQ}.

\subsection{EAirQ framework}
\label{subsec: EAirQ}

\begin{figure*}[ht!]
	\centering
	\includegraphics[width=0.8\textwidth]{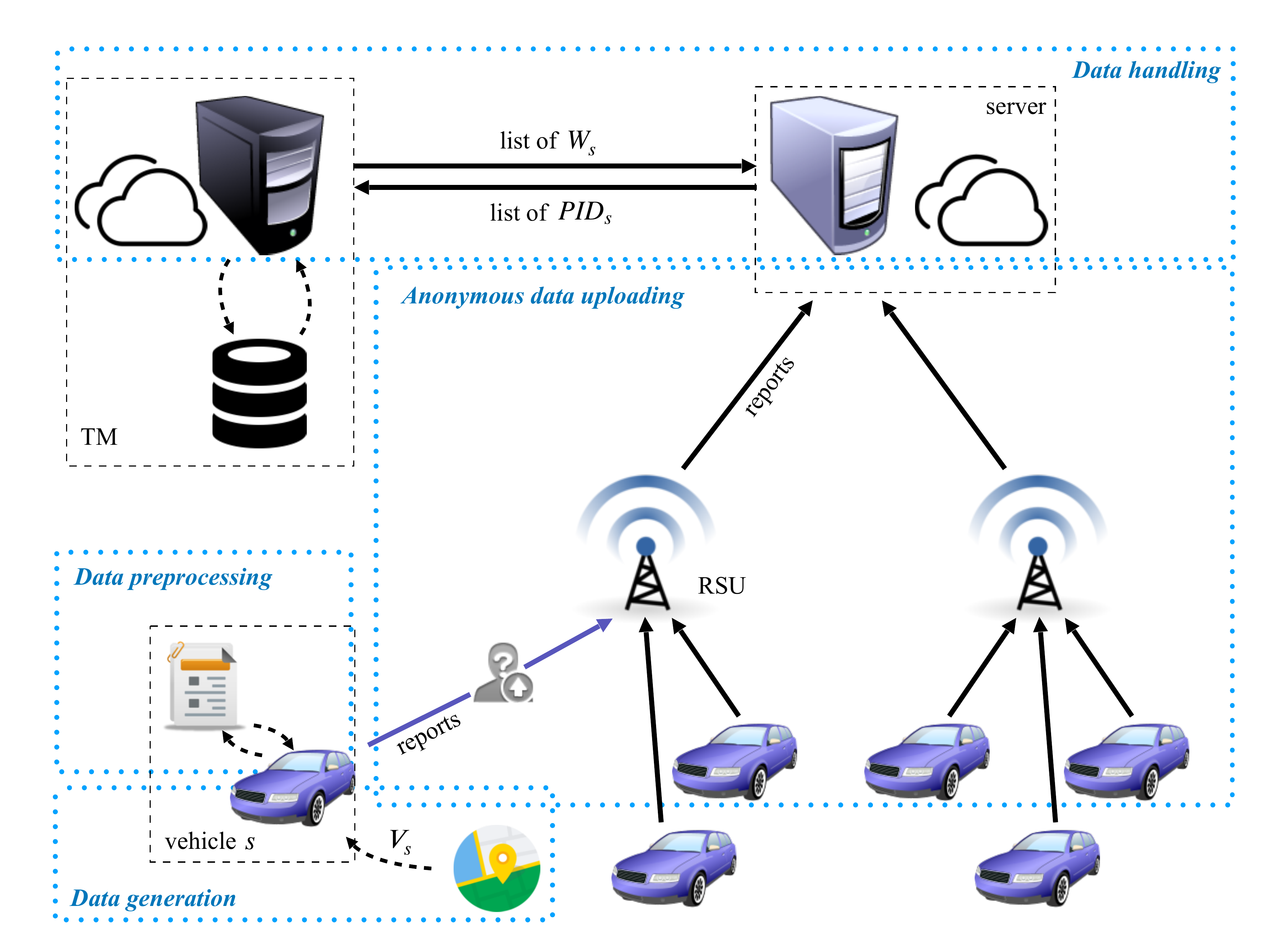}
	\caption{The EAirQ framework processes}
	\label{fig:system}
\end{figure*}

In EAirQ, there are four entities: vehicles (i.e., sources), RSUs, a server (i.e., the truth discovery server), and a Trusted Manager (TM). The TM not only acts as a TA introduced in Section~\ref{subsec:pre_anoy}, but also manages all vehicles. 
To be specific, it has but is not limited to the following functions: 1) the TM generates and distributes the system parameters including the ones necessary for anonymous authentication. 2) The TM can explore the real identity of a vehicle. 3) The TM maintains the historical weights for all vehicles. 4) The TM can update the weights for vehicles based on application-level user activities. For example, a user (i.e., a driver of a vehicle) who reads articles every day in the app for a long time is intuitively more reliable than a user who signed up several days before. 5) The TM works with cloud techniques and can communicate with the truth discovery server efficiently.

As shown in Figure~\ref{fig:system}, in each sensing cycle, a vehicle $s$ generates and perturbs the sensory data with the perturbation mechanism in the processes of \emph{data generation} and \emph{data preprocessing}, respectively. The perturbed data are expected to be sent to an RSU through anonymous communication in the \emph{anonymous data uploading} process. After collecting all the data from RSUs, the server performs the truth discovery task with the help of the TM in the last process named as \emph{data handling}. 
Now we describe the above four processes in details as follows:  

\textbf{Data generation}:
in this process, each vehicle $s$ generates observation values $V_s$ for the grids passed by. 
Besides, to get ready for the following processes, $s$ should set up an anonymous communication with RSUs, i.e., update the pseudo-ID, ${PID}_s$, for message signing and verification~\cite{moni2020scalable}. 
Note that the processes of system parameter distribution and anonymous communication establishment are not shown in Figure~\ref{fig:system} as they are not the main focuses of this work. 

\textbf{Data preprocessing}:
in this process, $s$ first masks $\{\theta_{s,j,g}\cdot v_{s,j}\mid  j\in\{1,2,\dots,c\}\}$, $\{\theta_{s,j,g}\cdot v_{s,j}^2\mid  j\in\{1,2,\dots,c\}\}$, and $\{\theta_{s,j,g}\mid  j\in\{1,2,\dots,c\}\}$ for each grid $g$ with the raw observation values. The detailed masking algorithm is given in Section~\ref{subsec_datamasking}. The masked values are denoted as $\beta_{1}^{s,j,g}$, $\beta_{2}^{s,j,g}$ and $\beta_{3}^{s,j,g}$. Then, $s$ performs the grid perturbation and value perturbation on the raw observation values, as described in Section~\ref{subsec:perturbation}. 

\textbf{Anonymous data uploading}: in each sensing cycle, $s$ uploads $\{\beta_{1}^{s,j,g}\mid j\in\{1,2,\dots,c\}\}$, $\{\beta_{2}^{s,j,g}\mid j\in\{1,2,\dots,c\}\}$, $\{\beta_{3}^{s,j,g}\mid j\in\{1,2,\dots,c\}\}$ and $\{\hat{v_{s,j}}\mid g^{s,j}=g\}$ for each grid $g\in G$. Recall that $\{\hat{v_{s,j}}\mid g^{s,j}=g\}$ is the perturbed data. The report is uploaded with the pseudo-ID ${PID}_s$. Thus, both the RSU and the server cannot link a report to a vehicle.

\textbf{Data handling}: 
the pseudo-IDs protect vehicles but bring a challenge to appending the estimated weight $w_{s,t}$ to the corresponding vehicle's record $W_s$ in each sensing cycle. Thus, after collecting all the reports from RSUs, the cloud server first requests $W_s$ for each $s$ from the TM by sending the list of ${PID}_s$.
The TM looks for the corresponding $W_s$ by exploring the true identity (i.e., the real ID ${RID}_s$) of $s$ with ${PID_s}$. One point worth mentioning is that, the TM only shares a list of $W_s$ without ${RID}_s$ to the server. 
Besides, the TM has the authority to update $W_s$ based on the application-layer user activities so that $W_s$ may change every sensing cycle.
Thus, the server cannot obtain the corresponding real identities of the vehicles from the TM in the process.  

After acquiring $W_s$ for all ${PID}_s$, the server estimates the weights and ground truths as follows: for grid $g$ with sufficient reports, the ground truth $v_{g,t}^*$ is estimated by the simplified truth discovery algorithm SST. For $g$ with insufficient reports, $v_{g,t}^*$ is estimated by the truth discovery algorithm ST. A threshold $\tau$ should be determined based on different scenarios. 
The final weight $w_{s,t}$ is calculated by averaging the two results of the two algorithms. The TM then appends the final weight to $W_s$ for $s$ by linking ${RID}_s$ with ${PID_s}$, which is not shown in Figure~\ref{fig:system}.

%% file: EAirQ_5_eva.tex
\section{Performance evaluation}
\label{sec:eva}
 \subsection{Truth discovery performance}
  In this section, we conduct simulations to evaluate the performance of truth discovery in AirQ and EAirQ. A common and widely accepted truth discovery algorithm introduced in~\cite{miao2019privacy} is simulated for comparison, denoted as \emph{TD}~(\textbf{T}ruth \textbf{D}iscovery). The simulation results and the performance comparison are given in Section~\ref{subsec:tdperformence_AirQ} and \ref{subsec:tdperformence_EAirQ}.
 
 \subsubsection{Simulation setup}
 \label{subsec:simset}
  We first introduce the simulation setup for evaluating the performance of truth discovery in this section.
  
 \textbf{Dataset of grids and truths}: we adopt a dataset containing the Air Quality Index~(AQI) from 34 base stations in January 2020 in Beijing, China~\cite{dataset}. Each base station in the dataset is regarded as a grid. The geographical distances among grids are calculated with the longitudes and latitudes of the base stations.
 The AQI values are used as original real truths. 
 We observe that there is nearly no temporal correlation because of the coarse granularity of record periods and grids.
 Thus, we interpolate three evenly spaced values between every two AQI values~(i.e., in every hour). As a result, the sensing cycle is 15 minutes and there are 2973 real truths in total for each grid. We use $\hat{v_g}$ to denote the real truth of grid $g$.
 
 \begin{figure}
 	\centering
	\begin{subfigure}{0.5\linewidth}
		\centering
		\includegraphics[width=\linewidth]{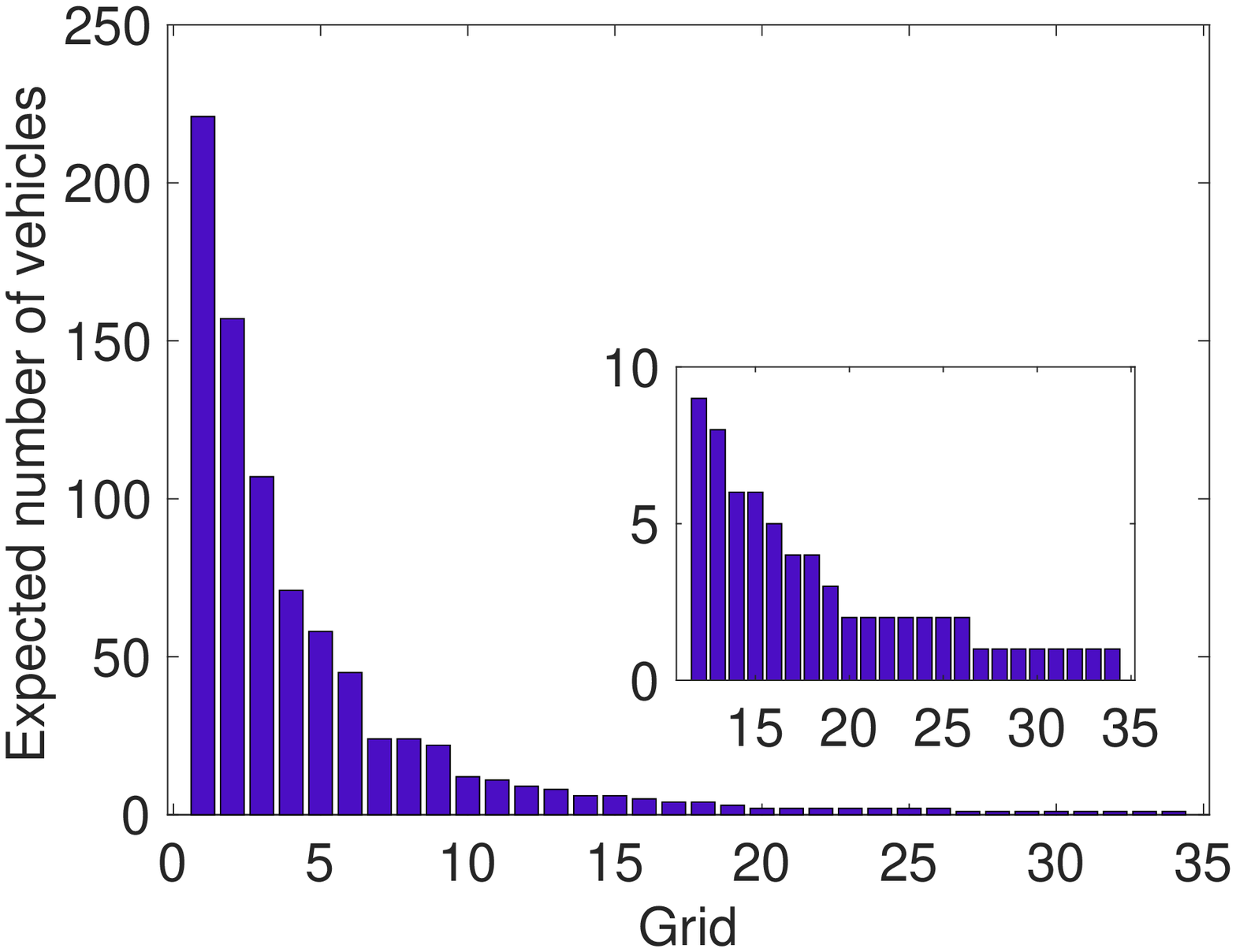}
		\caption{A Zipf distribution \\ example} \label{fig:zipf}
 	\end{subfigure}%
 	\begin{subfigure}{0.5\linewidth}
	 	\centering
	 	\includegraphics[width=\linewidth]{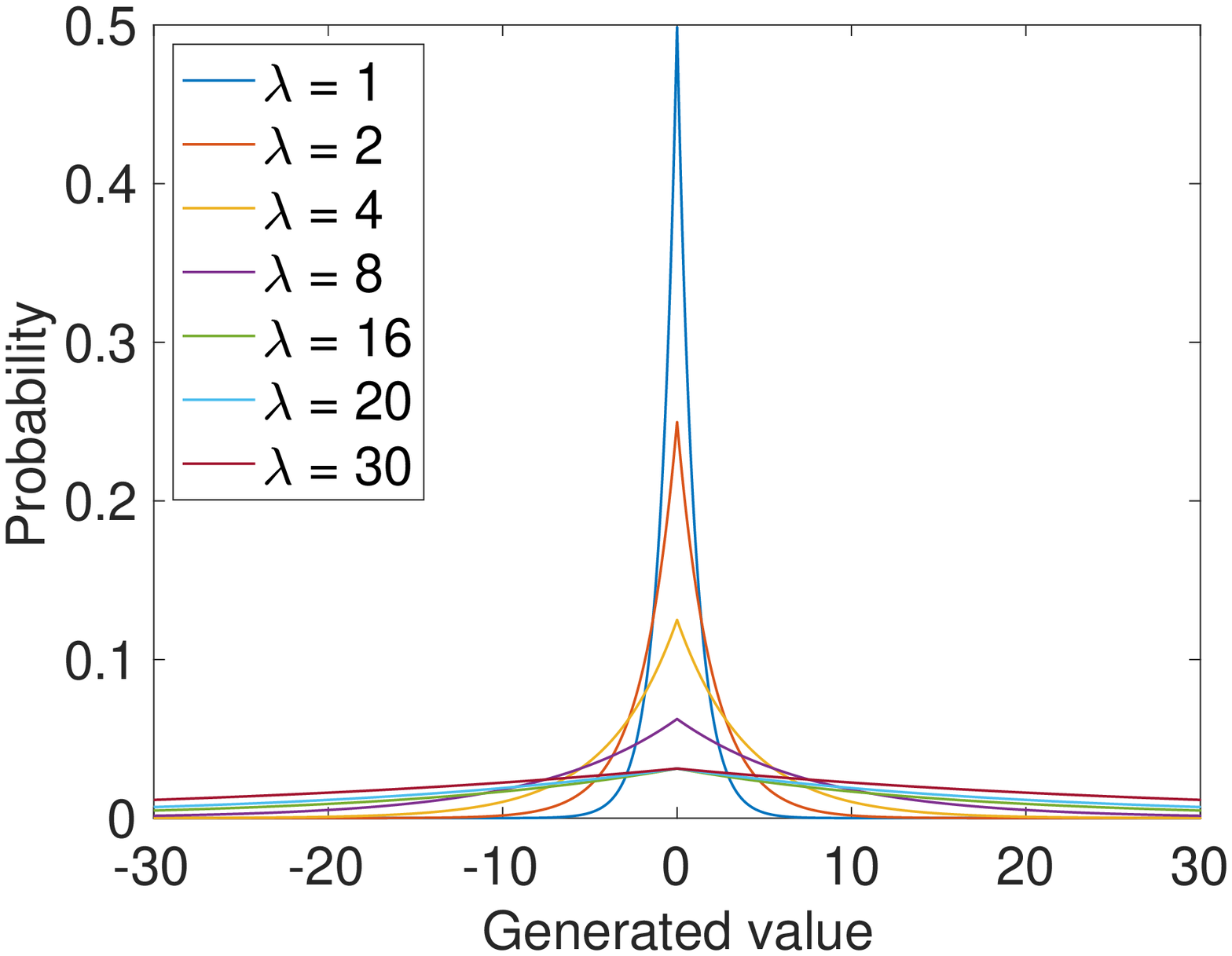}
	 	\caption{Laplace distributions\\ with different $\lambda$} \label{fig:lap}
    \end{subfigure}%
 	\caption{Distributions} 
 \end{figure}
 
\textbf{Long tail phenomenon}: in the dataset, some base stations~(e.g., Qianmen) are located in busy commercial centers while some~(e.g., Yungang) out of the Fourth Ring Road of Beijing, i.e., not in busy areas. This intuitively leads to an obvious difference in vehicle densities, which is similar to the observation in~\cite{xu2016understanding}, i.e., the long tail phenomenon. 
To simulate the phenomenon, we set the
expected number of vehicles passing by a grid following a Zipf distribution. In other words, in a sensing cycle, most of the vehicles are expected to pass by a small portion of grids in our simulations. One example is shown in Figure~\ref{fig:zipf}.
Note that we consider each sensing cycle independent so that the expected number is used as the mean value in a Poisson distribution to generate the exact number of vehicles passing by the grid in the cycle. 

\begin{figure}[ht]
	\centering
	\includegraphics[width=0.9\linewidth]{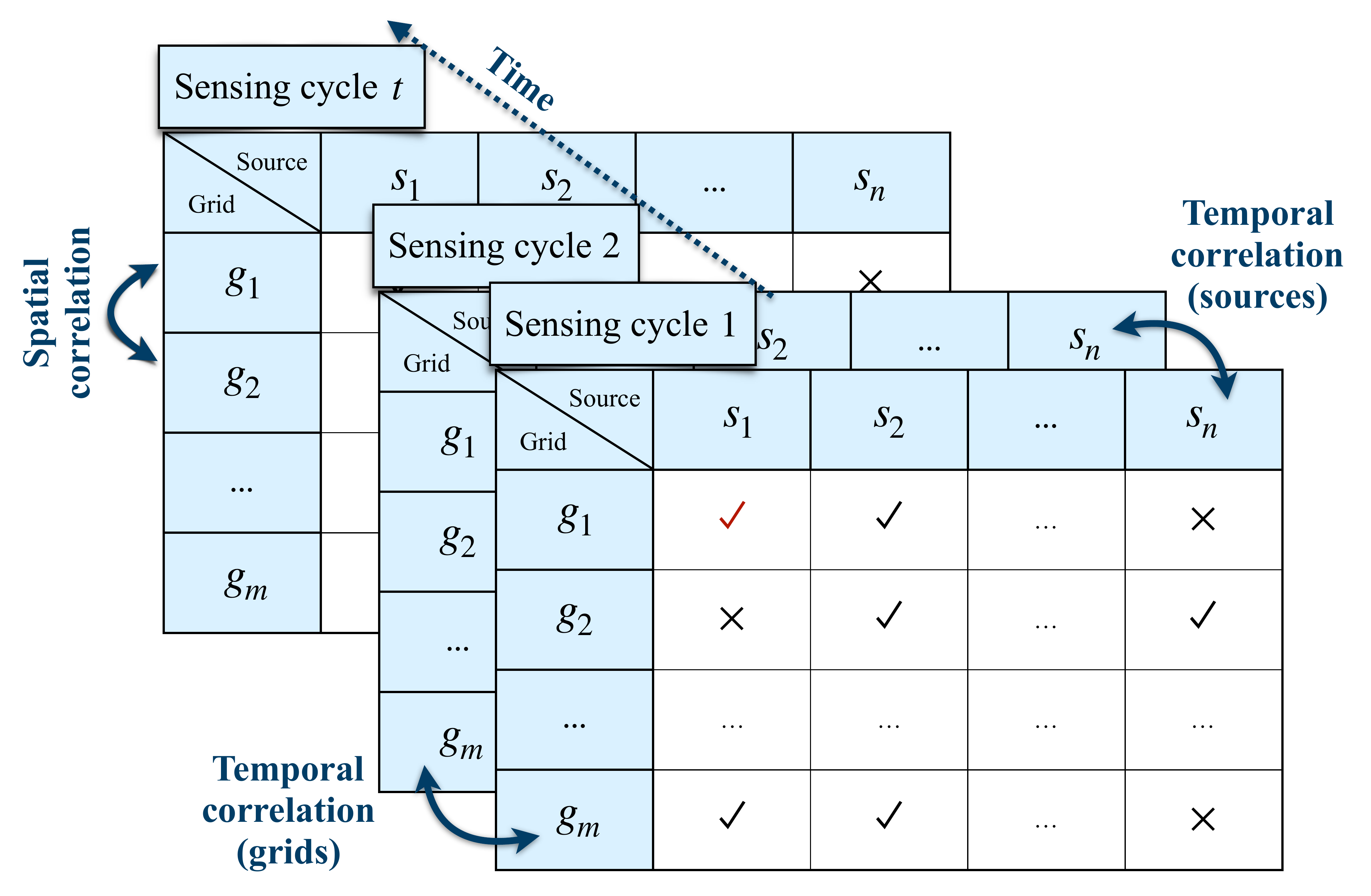}
	\caption{Observation values generation}
	\label{fig:ob}
\end{figure}

\textbf{Vehicle trajectories}: to simplify the simulation, instead of generating a real trajectory for a vehicle, we only set up a ``simplified trajectory" in advance, without considering the sequence of the vehicle movements. To be specific, in each sensing cycle, we randomly generate a list of vehicles for each grid $g$. The length of the list is the exact number of vehicles passing by $g$, i.e., the number generated above with the Poisson and Zipf distributions. An example is given in Fig.~\ref{fig:ob}. The red check mark represents $s_1$ passes by $g_1$ in the sensing cycle $1$ and contributes an observation value. $s_1$ also passes $g_m$ but does not pass $g_2$ in this cycle. 
We assume that each vehicle only contributes zero (cross mark in Fig.~\ref{fig:ob}) or one (check mark in Fig.~\ref{fig:ob}) observation value for one grid in each sensing cycle, as mentioned in Section~\ref{sebsec:prb_def}.

\textbf{Vehicle reliability}: considering the different precisions of onboard sensors and the possibility of malicious vehicles, a reliability value should be initialized for each vehicle in advance. 
In our simulations, we set a deviation value $\kappa_s$ following a truncated Normal distribution $\mathcal{TN}$ to represent the reliability.
Recall that the reliability of sources is unknown a priori in practice. The initialized reliability in the simulations is only used to generate corresponding observation values and never accessible to the server.

\textbf{Observation values}: as a systematic bias, $\kappa_s$ can be used as a multiplicative factor of each real truth $\hat{v_g}$ to generate the observation value for $s$. 
To be specific, $\hat{v_g}\cdot \kappa_s$ is the expected observation value for grid $g$ provided by $s$. We adopt a Normal distribution $\mathcal{N}$ to simulate the accidental bias. We choose the variance of the distribution as 0.2 so that the accidental bias has low impacts.
Supposing $g^{s,j}=g$, the observation value $v_{s,j}$ is generated from $\mathcal{N}(\hat{v_g}\cdot \kappa_s, 0.2)$. 

\textbf{Evaluation metrics}: we evaluate the performance of truth discovery based on two metrics.
\begin{itemize}
	\item \emph{Root-Mean-Square Error (RMSE)}: the average root-mean-square deviation of each sensing cycle. To make figures clear, we calculate the average RMSEs for each day, i.e., 96 sensing cycles.
	\item \emph{Valid Estimations}: the number of records, i.e., estimated ground truths, whose relative deviations from the real truths are less than a threshold.
\end{itemize}

\subsubsection{Performance of AirQ}
\label{subsec:tdperformence_AirQ}
To evaluate the truth discovery performance of AirQ, we first conduct simulations on 500 vehicles (considering the scenarios such as a residential area or a tourist town) and 2973 sensing cycles. In these simulations, two parameters warrant discussing, the source reliability and the threshold for valid estimations.

\textbf{Source reliability}:
the source reliability $\kappa_s\sim \mathcal{TN}(1.5,0.5,1,\sigma)$ where the parameters of the truncated Normal distribution are the upper limit, the lower limit, the mean and the standard deviation, respectively. Obviously, now all the observation values are in an acceptable range of $\lbrack0.5\hat{v_g},1.5\hat{v_g}\rbrack$. 
$\sigma$ are set differently to simulate the scenarios where most of the vehicles are normal~($\sigma=0.5$), a great number of vehicles are normal~($\sigma=1$), and a great number of vehicles are abnormal~($\sigma=2$). Note that although we use ``normal" and ``abnormal" to describe the sources with different reliabilities, all the sources with this setting are \emph{good}. In other words, these sources may have different $\kappa_s$ because of varying sensor precisions but the mean value is 1. It is reasonable and common in practice. 

\textbf{Threshold for valid estimations}: because there is no standardized threshold in related works, we intuitively set the thresholds as $15\%$, $20\%$ and $25\%$ for different error-tolerant levels. 
The reason is that the air pollution categories are defined by every 50 or 100 scores of AQI according to the Technical Regulation on Ambient Air Quality Index~\cite{AQIPDF}. For example, AQI among 0--50 represents the category of \emph{Excellent} and 200--300 represents \emph{Heavily polluted}. A deviation of $15\%$, $20\%$ or $25\%$ is acceptable considering the category step.

\begin{figure}[ht]
	\begin{subfigure}{0.5\linewidth}
		\centering
		\includegraphics[width=\linewidth]{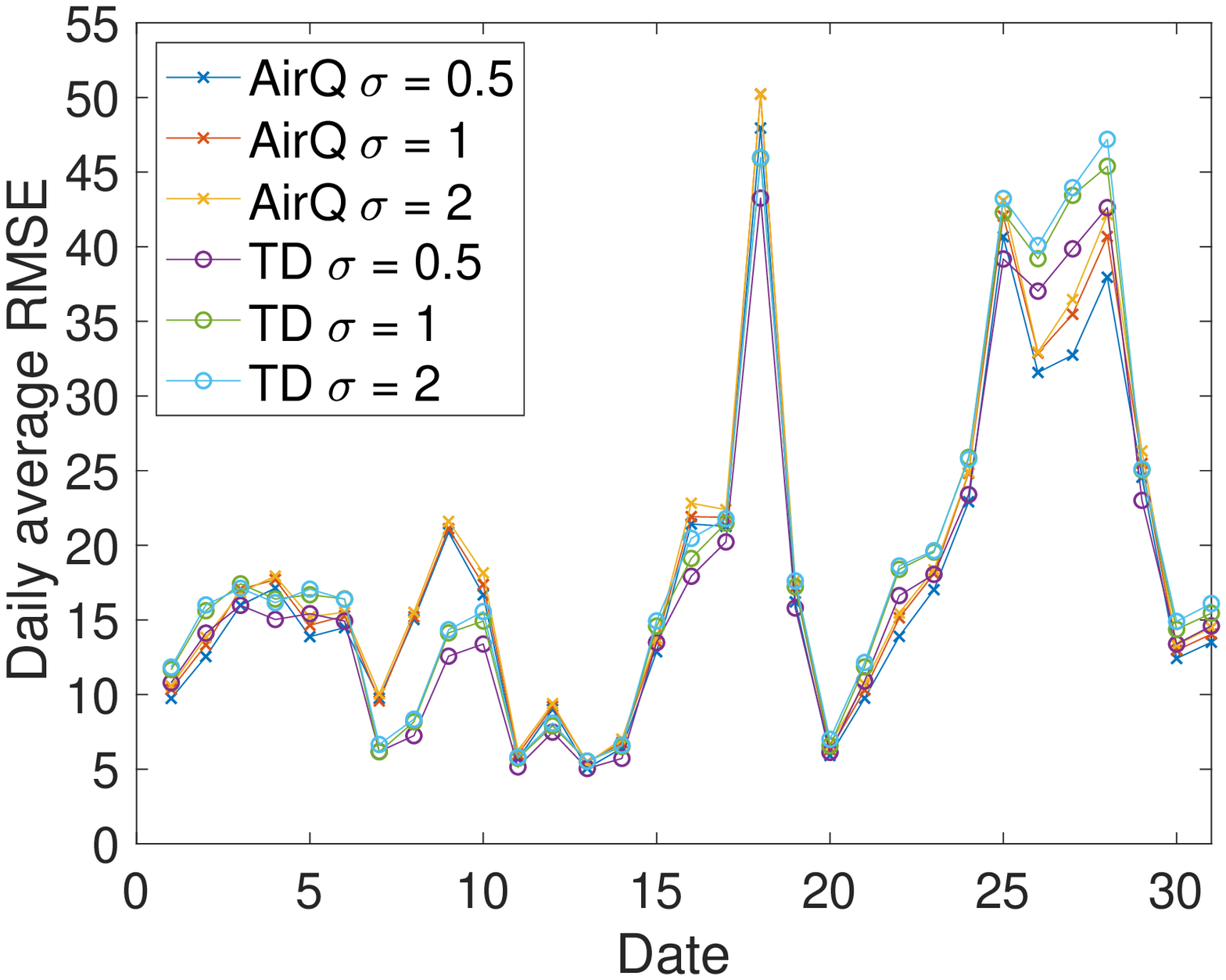}
		\caption{RMSEs\\~} \label{fig:perfor_a}
	\end{subfigure}%
	\begin{subfigure}{0.5\linewidth}
		\centering
		\includegraphics[width=\linewidth]{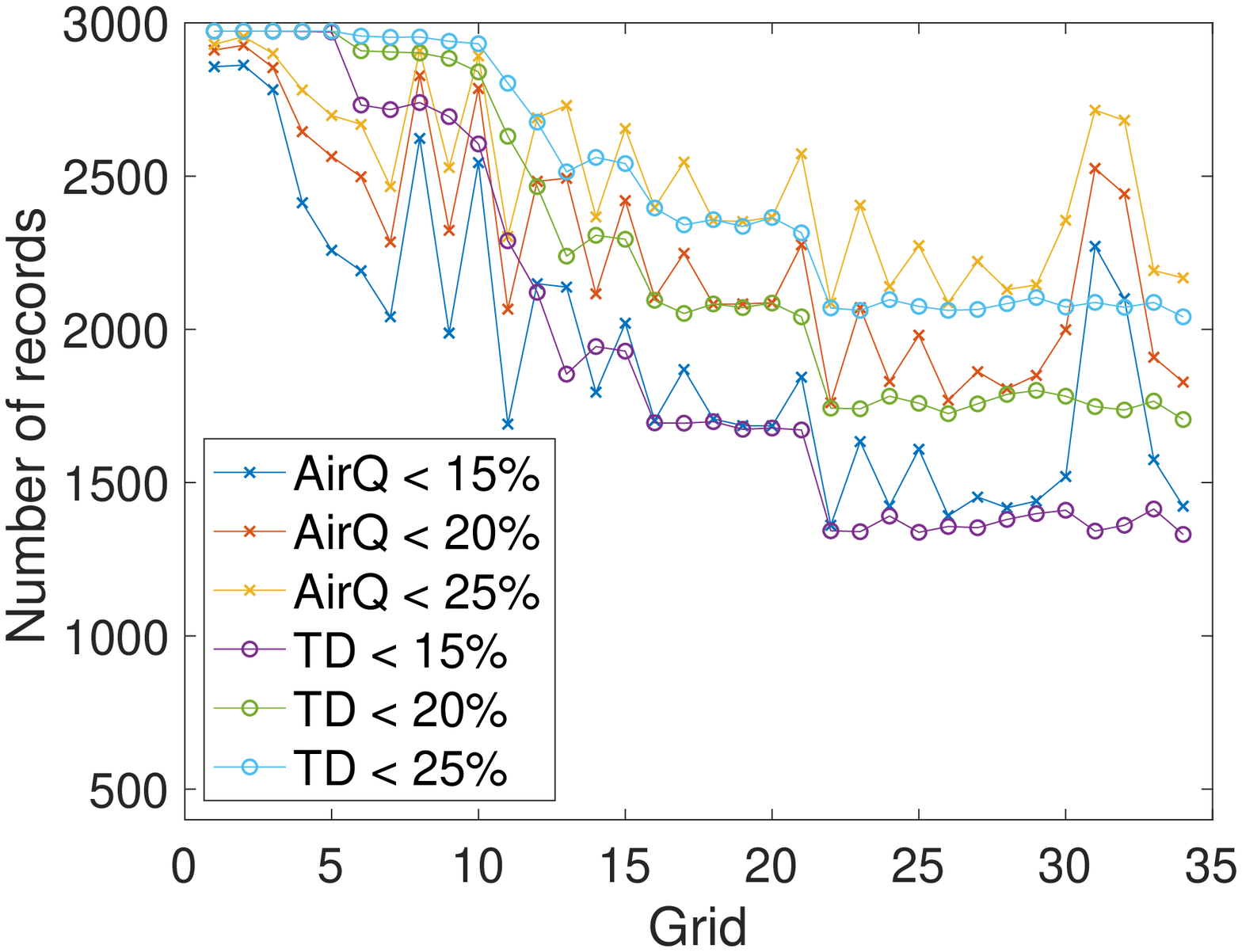}
		\caption{Valid estimations\\($\sigma = 0.5$)} \label{fig:perfor_b}
	\end{subfigure}%
	\newline
	\begin{subfigure}{0.5\linewidth}
		\centering
		\includegraphics[width=\linewidth]{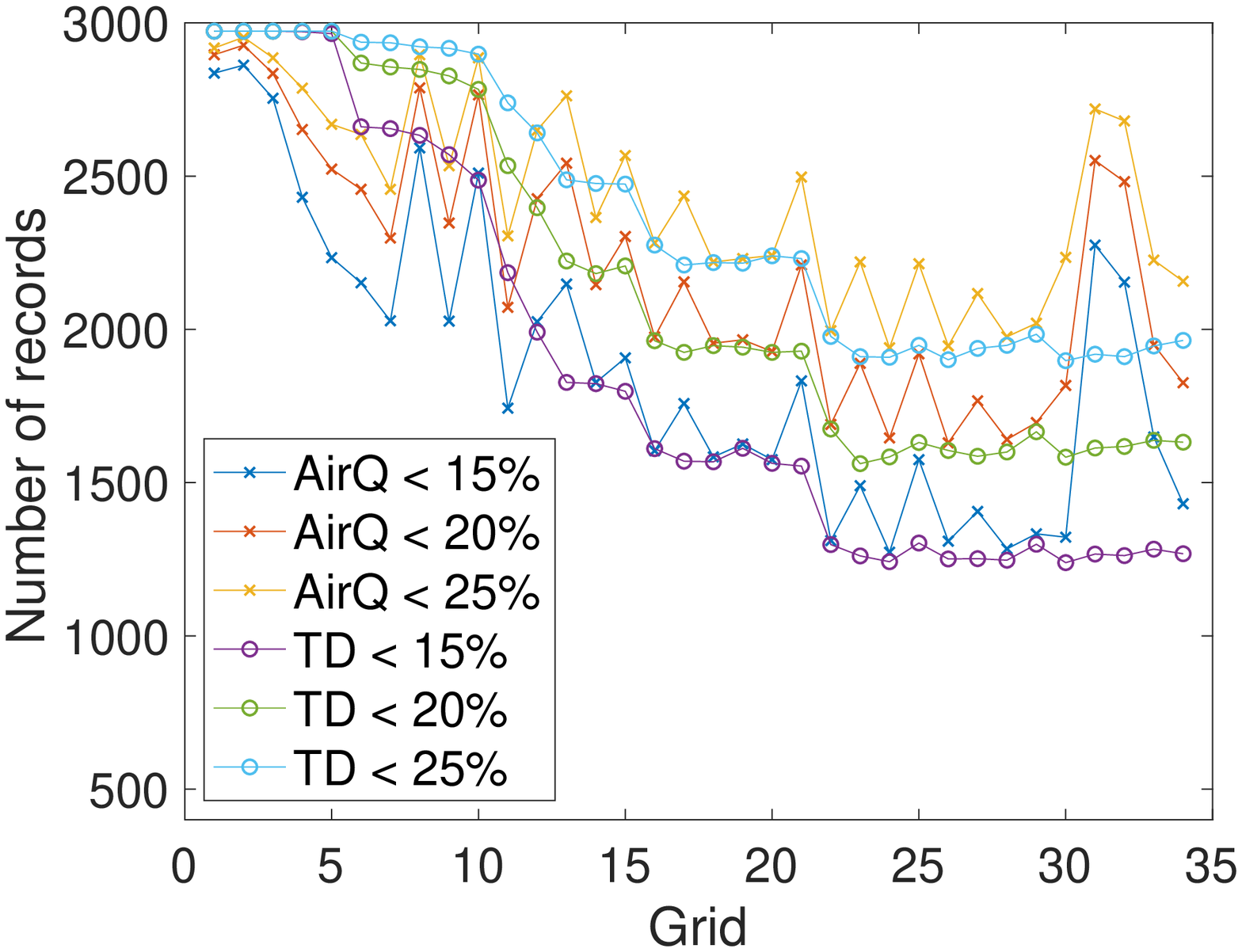}
		\caption{Valid estimations\\($\sigma = 1$)} \label{fig:perfor_c}
	\end{subfigure}%
	\begin{subfigure}{0.5\linewidth}
		\centering
		\includegraphics[width=\linewidth]{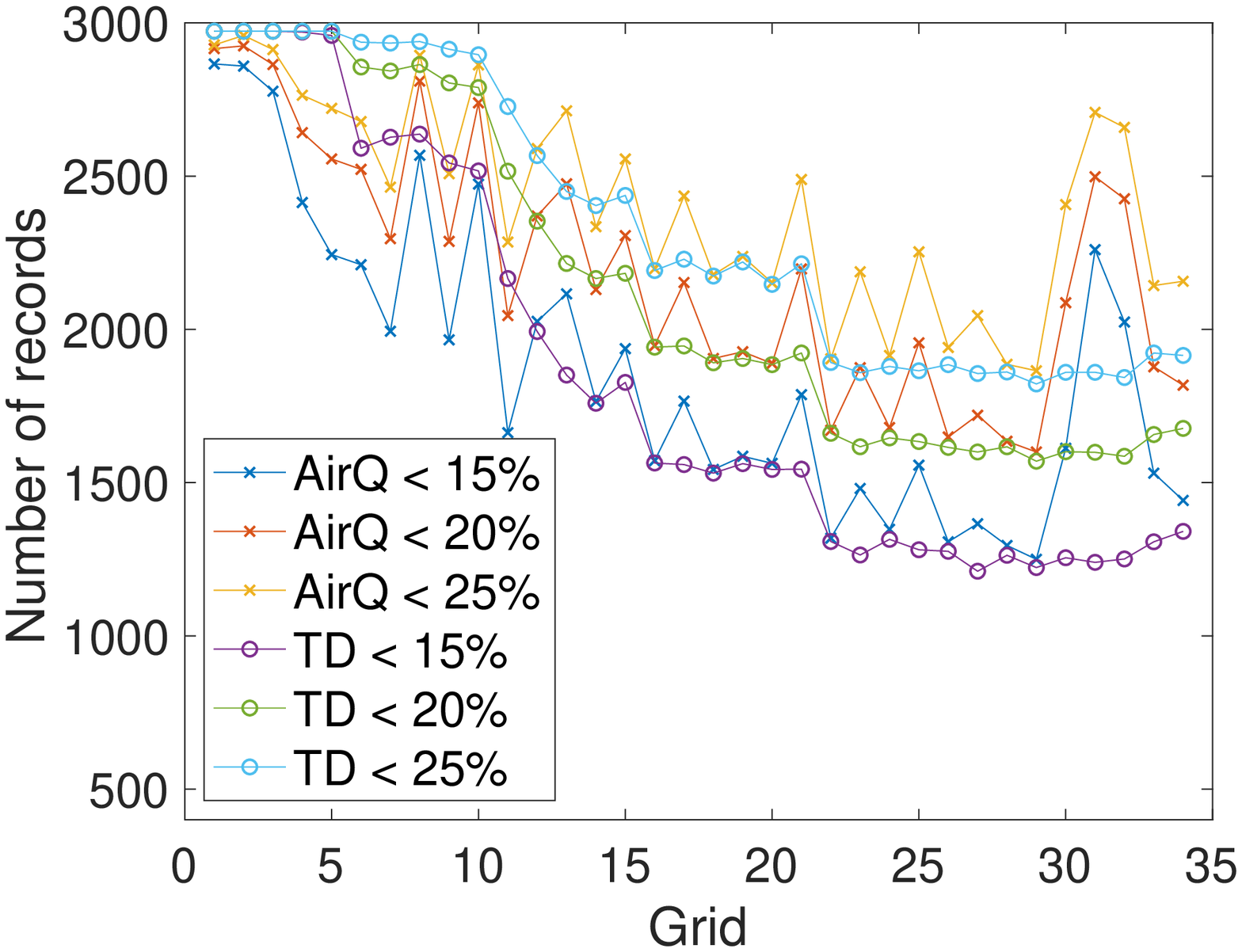}
		\caption{Valid estimations\\($\sigma = 2$)} \label{fig:perfor_d}
	\end{subfigure}%
	\caption{Performance of AirQ and TD with different $\sigma$}
	\label{fig:perfor}
\end{figure}
  
The simulation results are shown in Fig.~\ref{fig:perfor}. 
From Fig.~\ref{fig:perfor_a}, we can observe that in most of the cases, more normal vehicles lead to smaller RMSEs because more accurate observation values are provided. 
The performance of AirQ and TD is further compared in Fig.~\ref{fig:perfor_b},~\ref{fig:perfor_c}~and~\ref{fig:perfor_d}.
Obviously, AirQ performs better when there is not sufficient data~(i.e., grids 12 to 34). The results echo the aforementioned Zipf distribution. 
When there are sufficient data~(i.e., grids 1 to 11), the average of observation values is very close to the ground truth because the deviations of the sources follow a normal distribution with a mean of 1. 
However, the estimated results of AirQ incorporating the spatial and temporal correlations may deviate from the average values. Thus, TD works better for grids 1 to 11. 
This result is reasonable and motivated us to present EAirQ.

\begin{figure}
 	\begin{subfigure}{0.5\linewidth}
 		\centering
 		\includegraphics[width=\linewidth]{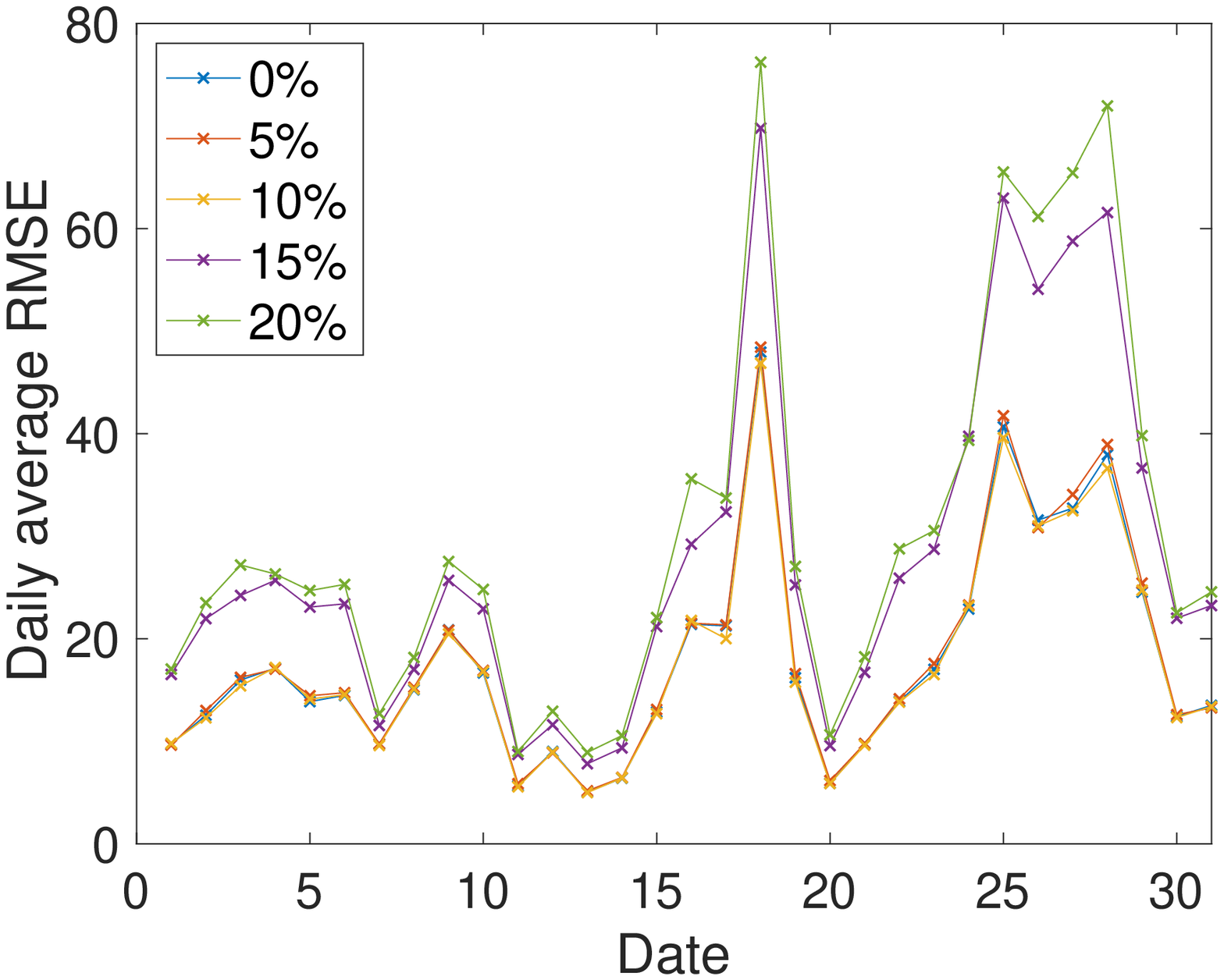}
 		\caption{RMSEs of AirQ} \label{fig:perfor3_a}
 	\end{subfigure}%
 	\begin{subfigure}{0.5\linewidth}
 		\centering
 		\includegraphics[width=\linewidth]{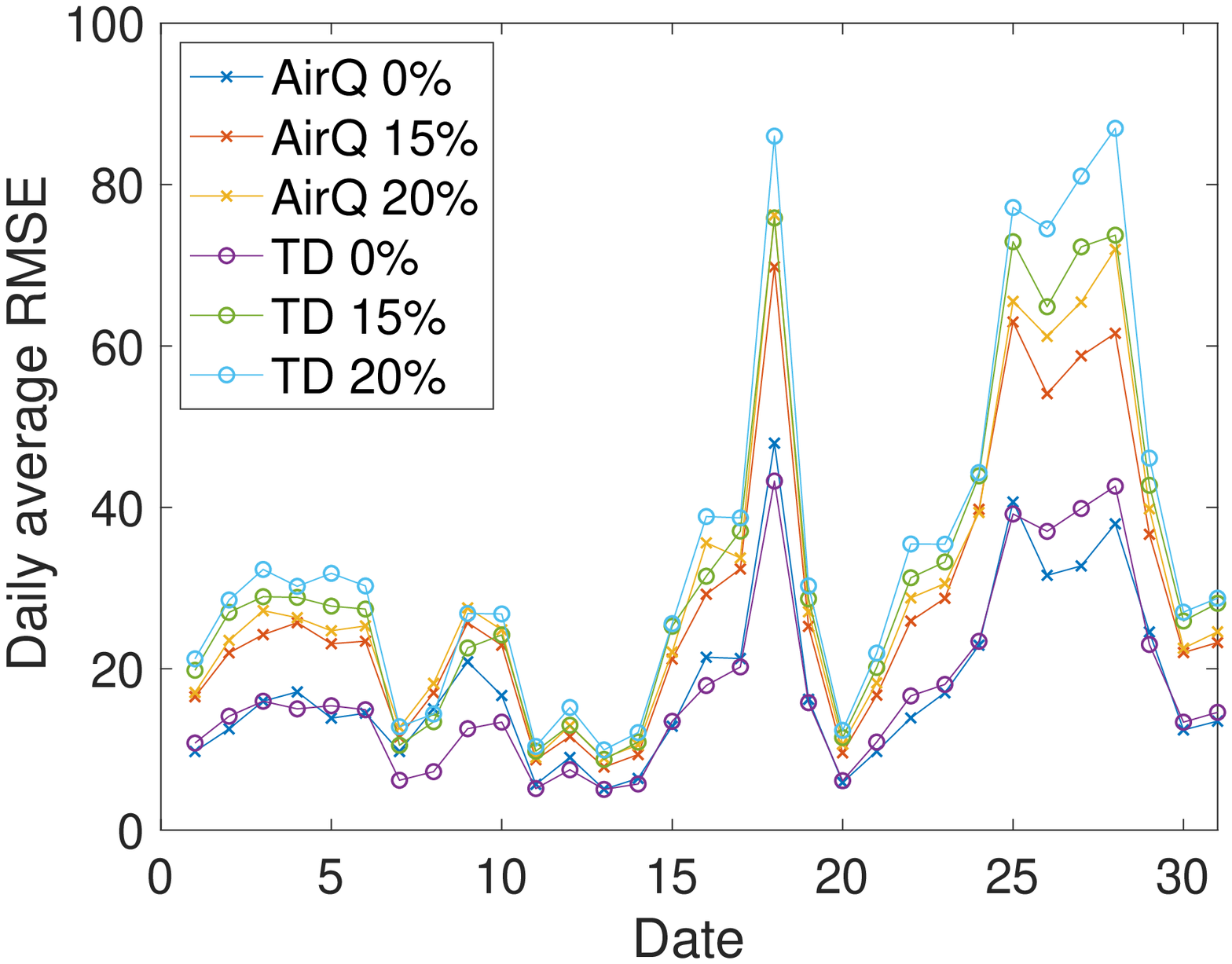}
 		\caption{RMSEs of AirQ and TD} \label{fig:perfor3_b}
 	\end{subfigure}%
 	\caption{Performance of AirQ and TD under the scenario with bad sources}
 	\label{fig:perfor3}
\end{figure}

Additionally, we conduct simulations under the scenario with \textit{bad} sources. Different from the good sources who have deviation values with a mean of 1, bad sources have a mean value significantly varying from 1. 
It can be caused by the same sensor defect or unfair competitions. For example, multiple bad sources are hired to work together for the sake of raising the reported air pollution of a tourist town. In the simulations, we set $\kappa_s\sim \mathcal{TN}(1.5,0.5,1,0.5)$ for good sources and $\kappa_s\sim \mathcal{TN}(2.5,1.5,2,0.5)$ for bad sources. The total number of sources is still 500 and the percentages of bad sources are $0\%$, $5\%$, $10\%$, $15\%$ and $20\%$. Results are shown in Figs.~\ref{fig:perfor3_a} and~\ref{fig:perfor3_b}.
We can observe that $15\%$ and $20\%$ bad sources result in higher RMSEs obviously. AirQ performs better than TD on these occasions.

\subsubsection{Truth discovery performance of EAirQ}
\label{subsec:tdperformence_EAirQ}
\begin{figure*}[hbt!]
	\centering
	\begin{subfigure}{0.32\linewidth}
		\centering
		\includegraphics[width=\linewidth]{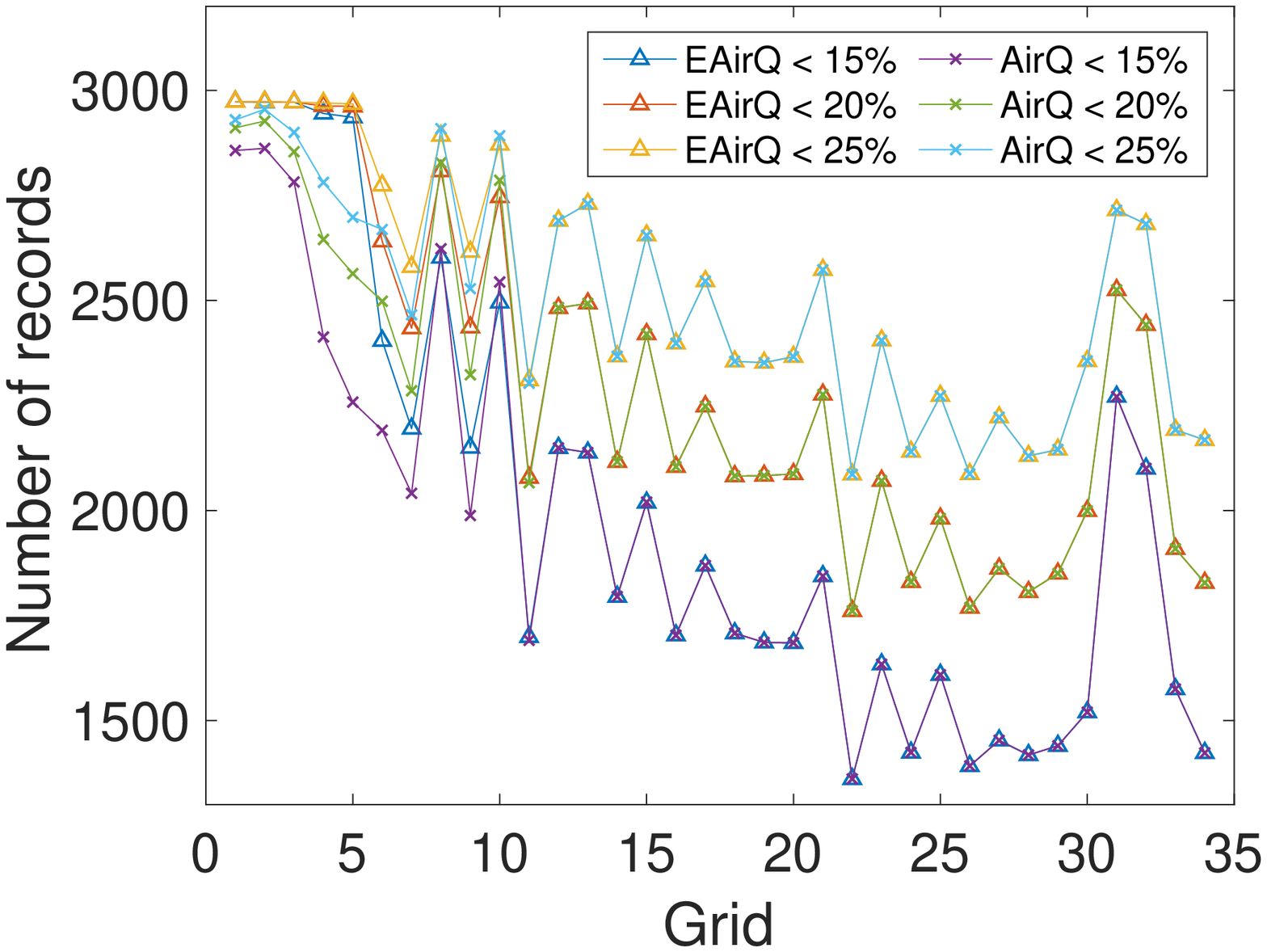}
		\caption{Valid estimations\\~} \label{fig:per_EAirQ_a}
	\end{subfigure}%
	\begin{subfigure}{0.32\linewidth}
		\centering
		\includegraphics[width=\linewidth]{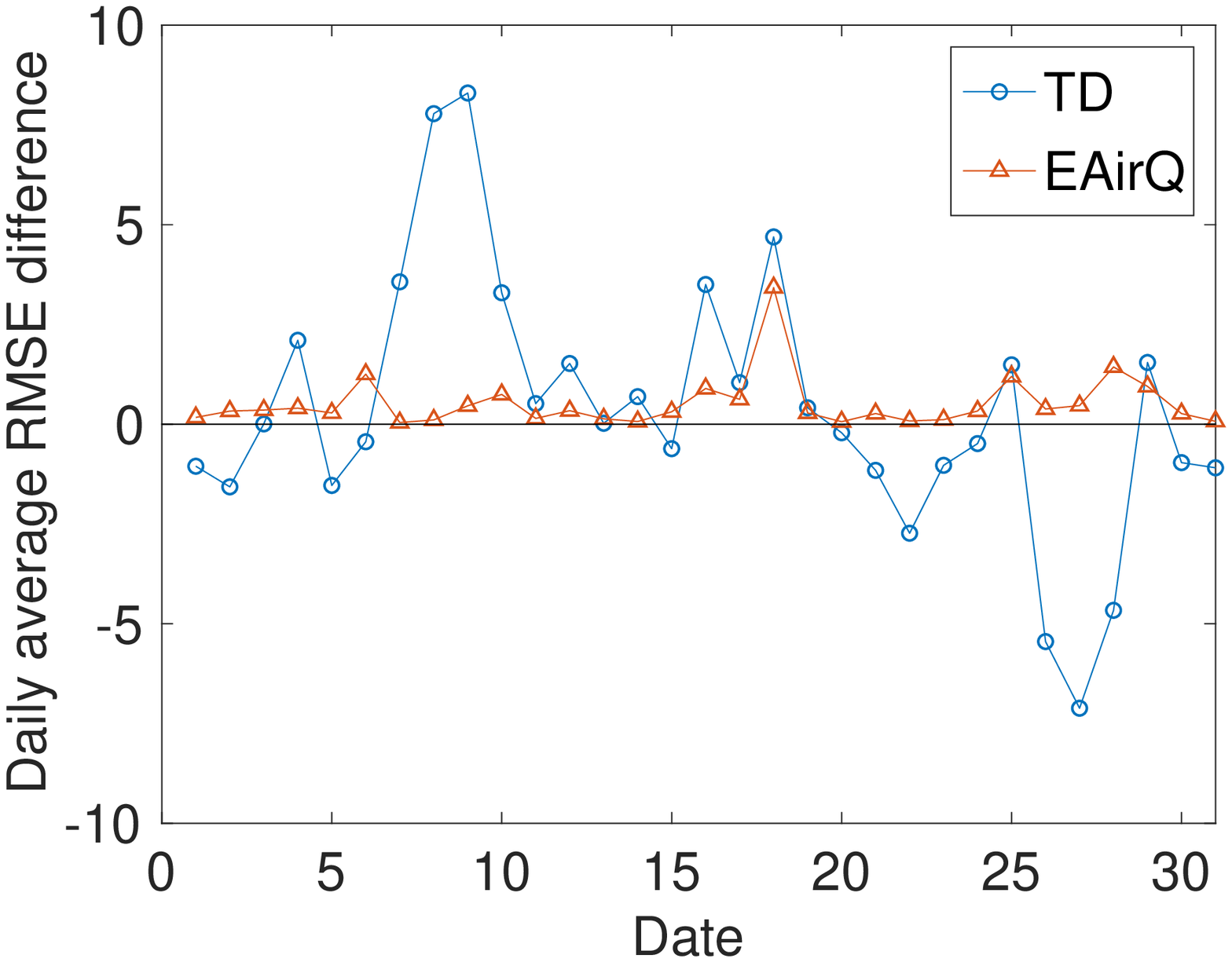}
		\caption{RMSE differences \\between TD/EAirQ and AirQ} \label{fig:per_EAirQ_b}
	\end{subfigure}%
	\begin{subfigure}{0.32\linewidth}
		\centering
		\includegraphics[width=\linewidth]{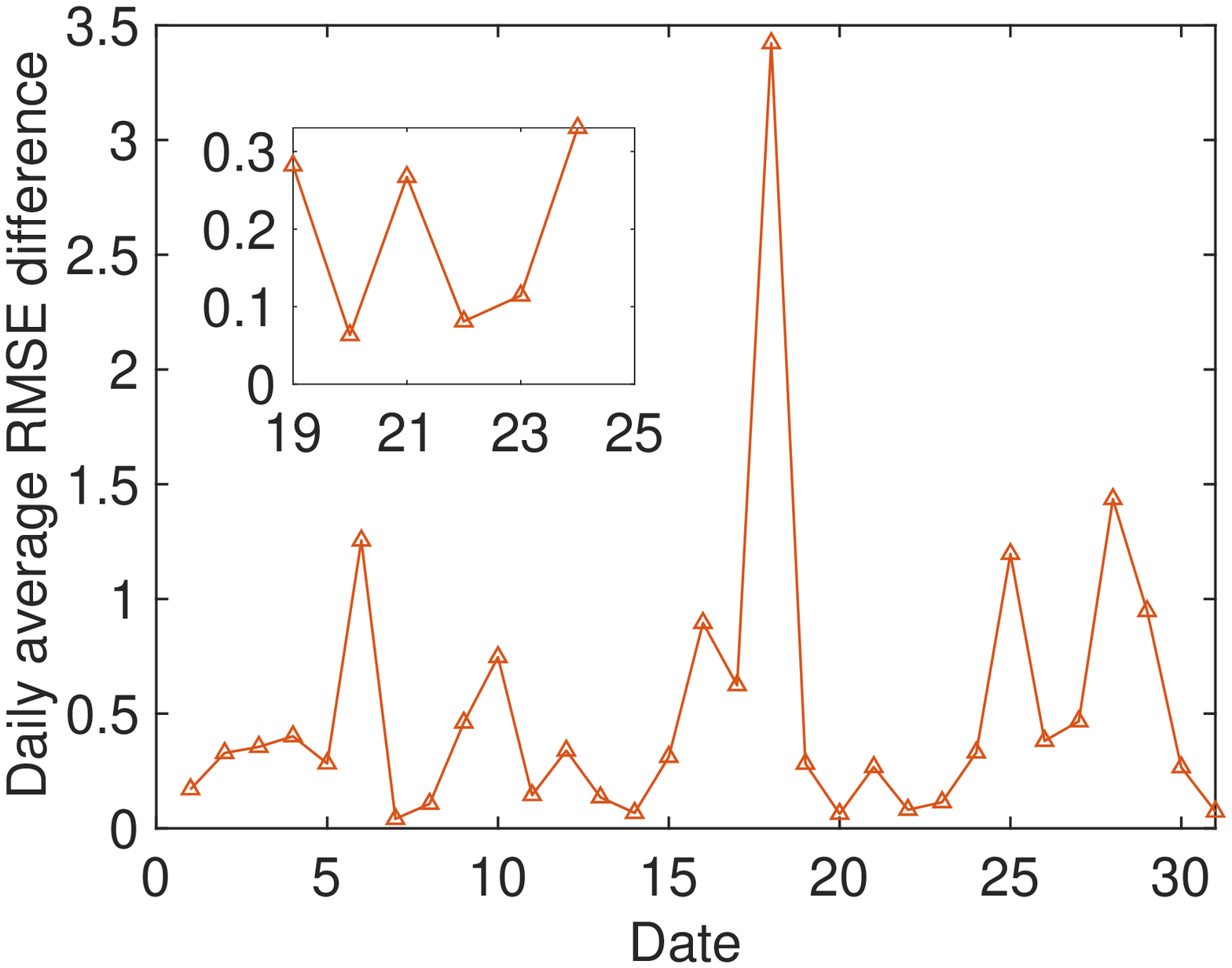}
		\caption{RMSE differences \\between EAirQ and AirQ} \label{fig:per_EAirQ_c}
	\end{subfigure}%
	\caption{Performance of EAirQ with good sources}
	\label{fig:per_EAirQ}
\end{figure*}

\begin{figure*}[hbt!]
	\centering
	\begin{subfigure}{0.32\linewidth}
		\centering
		\includegraphics[width=\linewidth]{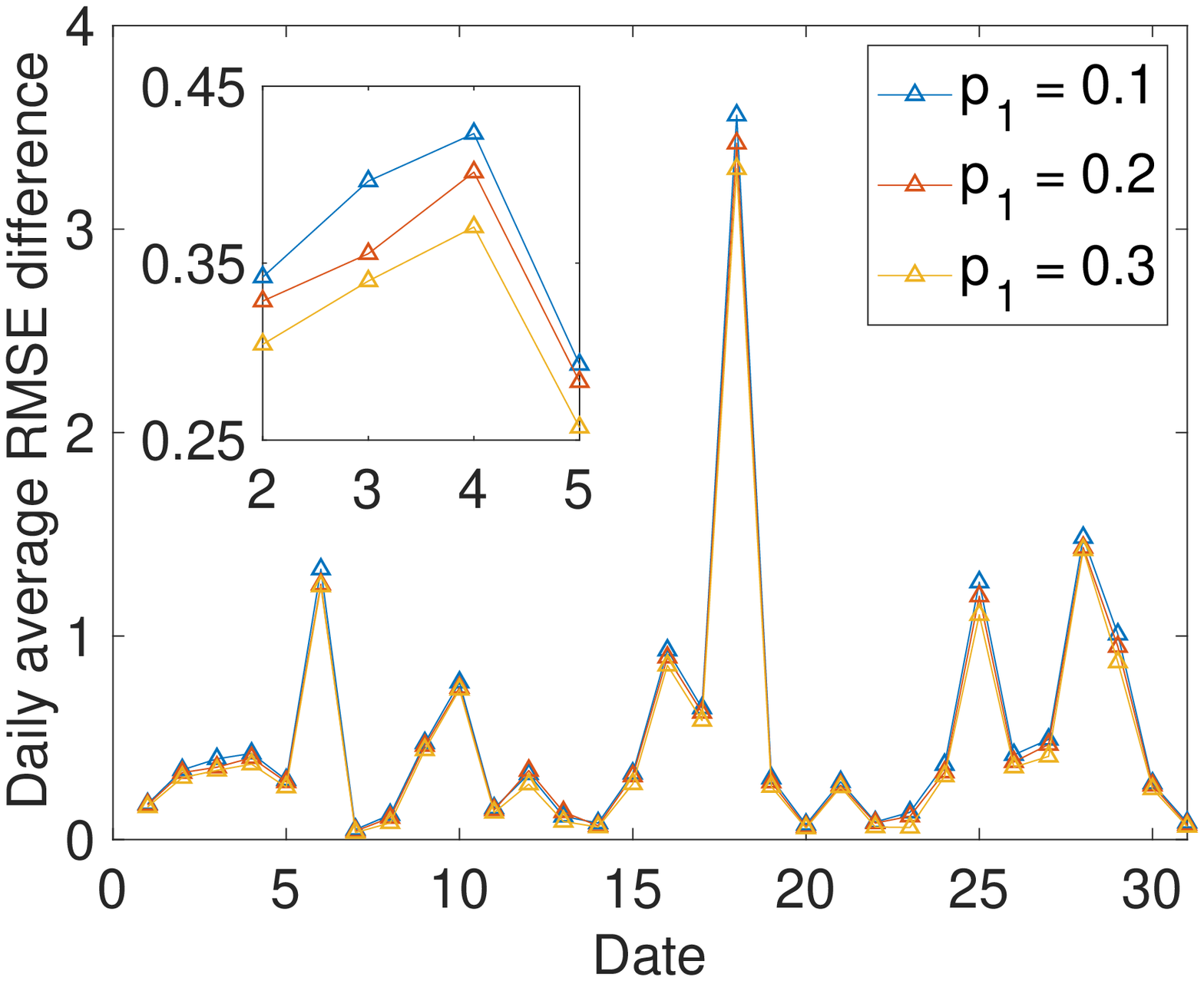}
		\caption{Different $p_1$} \label{fig:per_EAirQ_para_a}
	\end{subfigure}%
	\begin{subfigure}{0.32\linewidth}
		\centering
		\includegraphics[width=\linewidth]{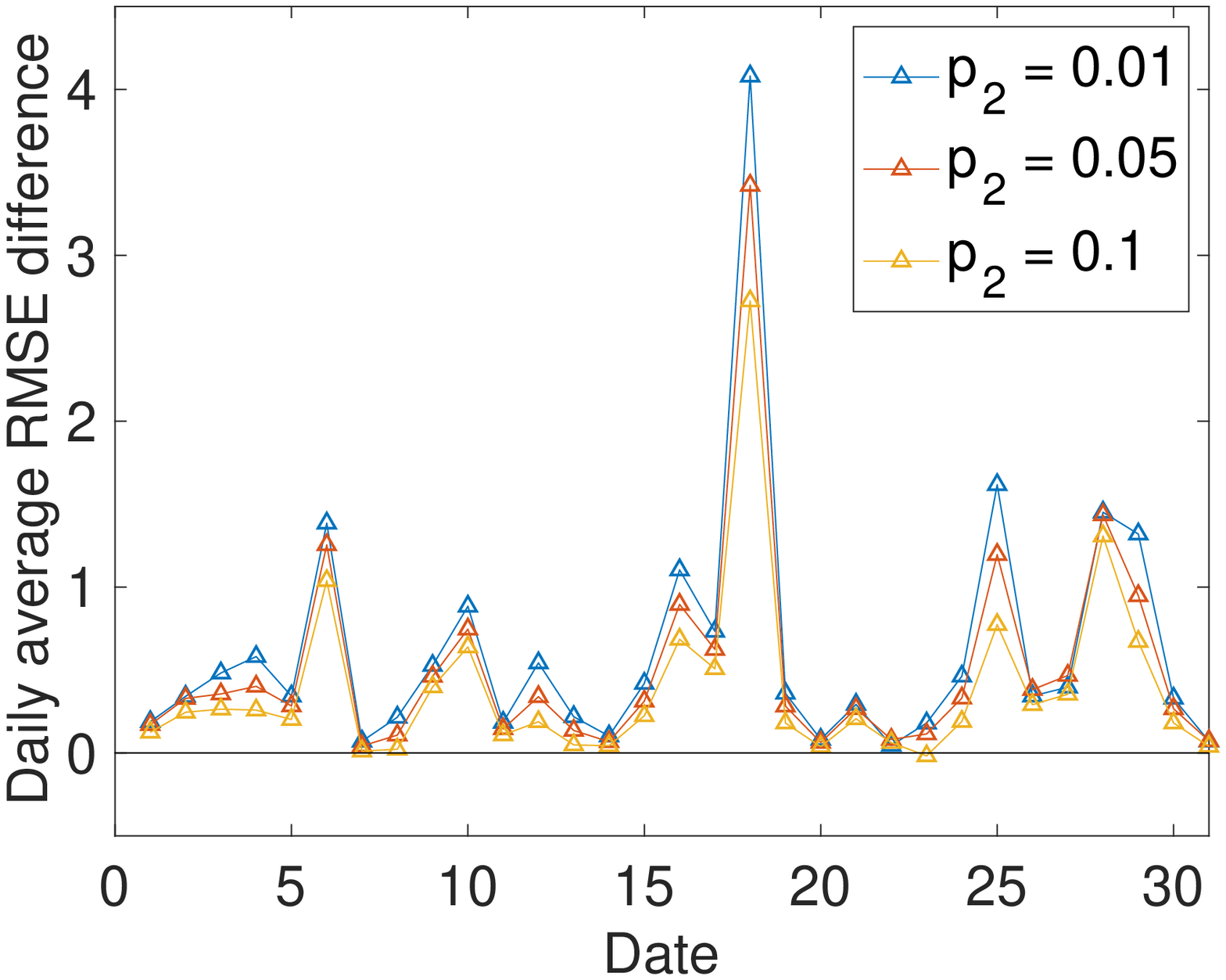}
		\caption{Different $p_2$} \label{fig:per_EAirQ_para_b}
	\end{subfigure}%
	\begin{subfigure}{0.32\linewidth}
		\centering
		\includegraphics[width=\linewidth]{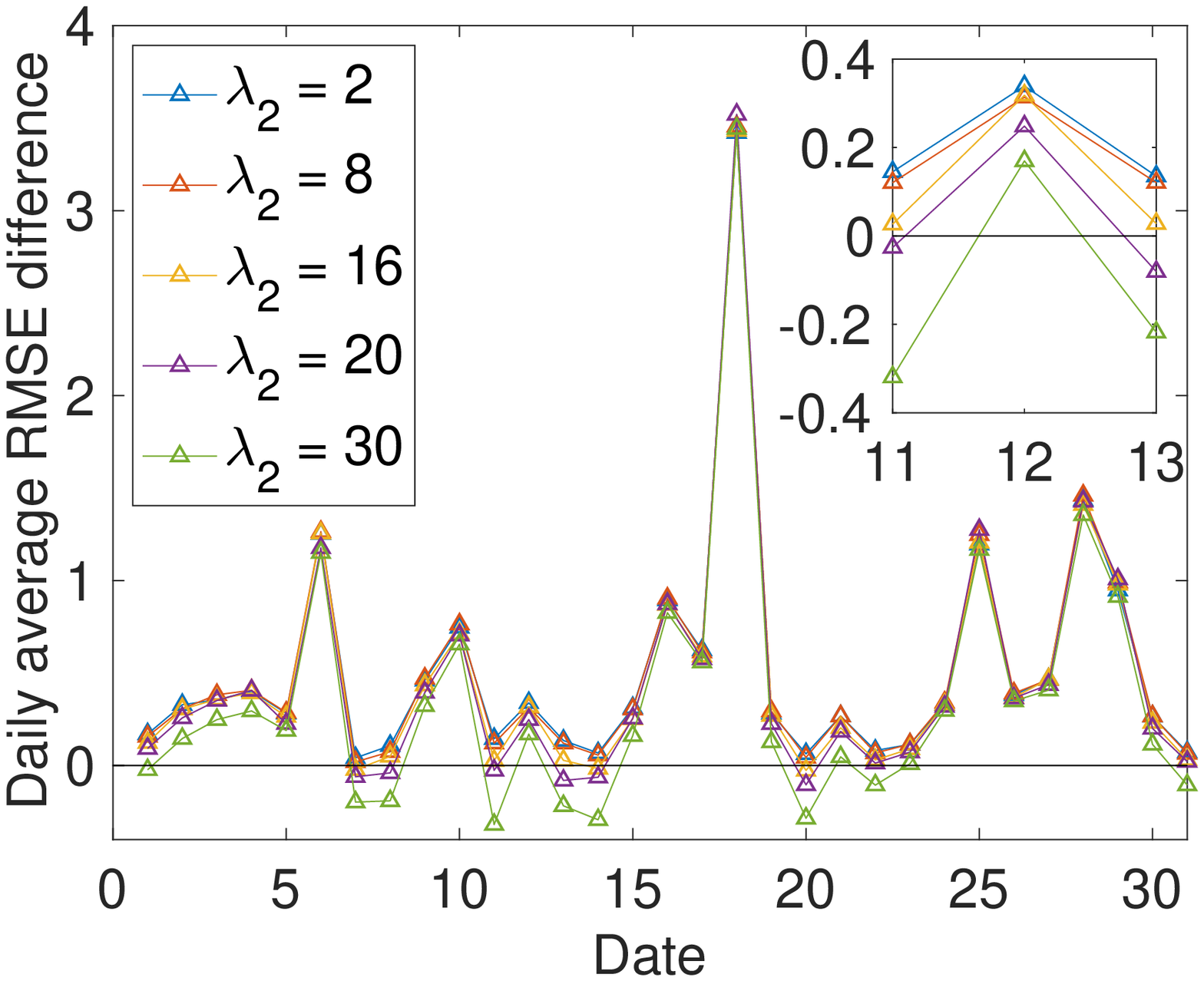}
		\caption{Different $\lambda_2$} \label{fig:per_EAirQ_para_c}
	\end{subfigure}%
	\caption{RMSE differences between EAirQ and AirQ with different perturbation parameters}
	\label{fig:per_EAirQ_para}
\end{figure*}

Because of the aforementioned limitation of AirQ, we further propose EAirQ and conduct simulations. In this section, we first discuss the newly-introduced parameters. Then we compare the performance of TD, AirQ and EAirQ with good sources. In addition, we analyze the impact of the perturbation mechanism on the accuracy of truth finding. Finally, we compare the performance of TD, AirQ and EAirQ with bad sources. 

We introduced five new parameters in the simulations:
\begin{itemize}
	\item $p_1$: the probability with which a vehicle removes an observation value from the report in the grid perturbation process, as defined in Equation~\eqref{equ_rand_respon}. 
	Recall that we generate a list of vehicles for each grid based on a Zipf distribution. We observed that, under the simulation settings, the total number of observation values provided by a vehicle (i.e., the number of grids passed by) in a sensing cycle usually has a mean less than 10. It corresponds with the practical situation: a vehicle usually cannot travel the majority of grids in one cycle. With this observation, one acceptable value of $p_1$ is 0.2. Then the expected number of removed observation values for a vehicle in a sensing cycle is less than 2.
	\item $p_2$: the probability with which a vehicle generates an observation value for a grid it does not pass by in one sensing cycle in the grid perturbation process. As mentioned in Section~\ref{subsec:perturbation}, we suggest setting a small value for $p_2$ considering the accuracy of truth discovery. For example, if $p_2 = 0.05$, recall that there are 31 grids in total and the number of grids a vehicle $s$ passes by is usually less than 10, and the expected number of simulated observation values is more than $(31-10)\times 0.05 =1.05$ for $s$. 
	\item $\lambda_1$: the scale parameter for the Laplace distribution $\mathcal{L}(0, \lambda_1)$ which is used to generate $\psi_1$ in the grid perturbation process. Because the AQI usually ranges from 20 to 100 in the dataset we adopt, adding a single-digit noise is acceptable. The probability density functions of the Laplace distributions with different scale parameters are shown in Fig.~\ref{fig:lap}.
	Thus, intuitively, $\lambda_1$ is better to be set around 2. 
	\item $\lambda_2$: the scale parameter for the Laplace distribution $\mathcal{L}(0, \lambda_2)$, which is used to generate $\psi_2$ in the value perturbation process. Similar to $\lambda_1$, setting $\lambda_1$ to around 2 is reasonable. Because every value perturbed with $\lambda_1$ is expected to be perturbed again with $\lambda_2$, we suggest setting $\lambda_1$ smaller than $\lambda_2$. Intuitively, we set $\lambda_1 = 1.5$ and $\lambda_2 = 2$ in our simulation.
	\item $\tau$: the threshold number of reports for a grid in a sensing cycle, which is used in the data handling process. 
	Considering both the simulation results analyzed in Section~\ref{subsec:tdperformence_AirQ} and the Zipf distribution shown in Fig~\ref{fig:zipf}, we set $\tau$ to 10, which is approximately the dividing line between grids 1 to 11 and 12 to 34. 
\end{itemize}

We first conduct simulations on 500 good vehicles with $\sigma = 0.5$, $p_1 = 0.2$, $p_2 = 0.05$, $\lambda_1 = 1.5$, $\lambda_2 = 2$ and $\tau = 10$. The valid estimations are shown in Fig.~\ref{fig:per_EAirQ_a}. 
Obviously, AirQ and EAirQ perform nearly the same for grid 12 to 34, which echoes the design of EAirQ, i.e., to use the ST algorithm for grids with insufficient data. What we concern about is grids 1 to 11. It shows that EAirQ has more valid estimations than AirQ, which indicates that the goal of EAirQ is achieved. 

To get clearer observations, we introduce a new evaluation metric based on the daily RMSE: \emph{daily RMSE difference}. It is the difference between two daily RMSEs. In our simulations, we always use the RMSEs of AirQ to subtract that of TD or EAirQ. Thus, if the difference is larger than 0, we can say TD or EAirQ works better than AirQ. The comparison results are shown in Fig.~\ref{fig:per_EAirQ_b} and~\ref{fig:per_EAirQ_c}. It is clear that EAirQ always has a positive difference value, which shows that EAirQ works better than AirQ.

 \begin{table}[h]
	\begin{center}
		\renewcommand{\arraystretch}{1.3}
		\centering
		\caption{Parameter settings for comparison}\label{tab:setting_para}
		\begin{tabular}{|c|c|c|c|c|}
			\hline
		    \multirow{2}{*}{Results }& \multicolumn{4}{c|}{Parameters}  \\\cline{2-5}
			 & $p_1$ & $p_2$ & $\lambda_1$ & $\lambda_2$ \\ \hline \hline
			Fig.~\ref{fig:per_EAirQ_para_a} & \tabincell{c}{0.1, 0.2 \\or 0.3}& 0.05& 1.5&2  \\ \hline
			Fig.~\ref{fig:per_EAirQ_para_b} & 0.2& \tabincell{c}{0.01, 0.05 \\ or 0.1}& 1.5&2  \\ \hline
			Fig.~\ref{fig:per_EAirQ_para_c} & 0.2& 0.05& 1.5&\tabincell{c}{2, 8, 16, \\ 20 or 30}  \\ \hline
		\end{tabular}
	\end{center}
\end{table}

Fig.~\ref{fig:per_EAirQ_para} shows the impact of different perturbation parameters. 
The settings of parameters are listed in Table~\ref{tab:setting_para}. 
Note that because both $\lambda_1$ and $\lambda_2$ are used to generate the Laplace noise and $\lambda_2$ affects much more data than $\lambda_1$, we only simulate with a different $\lambda_2$ as an example.
We can observe that: 1) a smaller $p_1$ or $p_2$ results in a higher RMSE difference. In other words, the estimated ground truths are more accurate. 2) The impact of $p_2$ is more significant than $p_1$ because a larger $p_2$ introduces more imitated reports. When $p_2 = 0.1$, EAirQ even performs a bit worse than AirQ on the 23rd day. 3) A higher $\lambda_2$ leads to worse performance. 
Negative difference values are observed if $\lambda_2 > 16$.
 These results are reasonable because a higher $p_1$, $p_2$ or $\lambda_2$ leads to more noise on the reports. Controlling the bias in an acceptable range and finding a balance between privacy and accuracy can be achieved by adjusting the parameters.
Besides, an interesting observation in Fig.~\ref{fig:per_EAirQ_para} is that, the results with $\lambda_2=2$ and $\lambda_2=8$ are close to each other. It is because we only use the perturbed data when there are sufficient reports and the symmetric Laplace noise can be canceled to some extent. 
In practice, $\lambda_2$ can be set a bit larger than 2 but the issues of weight management should be considered.
We discuss more in Section~\ref{sec:dis}.

\begin{figure}[ht]
	\centering
	\includegraphics[width=0.9\linewidth]{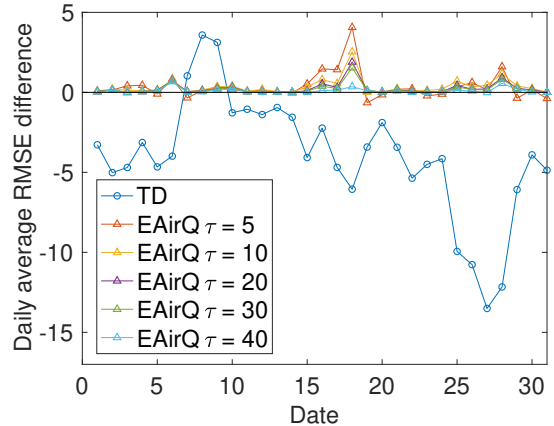}
	\caption{Performance of EAirQ with 15\% bad sources}
	\label{fig:per_EAirQ_bad}
\end{figure}

Based on what we observe from Fig.~\ref{fig:perfor3}, 0\% to 10\% bad sources do not make much difference under the simulation settings. Thus, we conduct simulations under the scenario with $15\%$ bad sources to evaluate the performance of EAirQ further. 
The simulations still use the setting that $\sigma = 0.5$, $p_1 = 0.2$, $p_2 = 0.05$, $\lambda_1 = 1.5$, $\lambda_2 = 2$ and $\tau = 10$. 
Results are shown in Fig.~\ref{fig:per_EAirQ_bad}. It can be seen that EAirQ has a higher accuracy than TD and AirQ. One point worth mentioning is that the choice of threshold $\tau$ is important but complex. To be specific, the relationship between $\tau$ and the truth discovery accuracy is not monotonically increasing or decreasing.
For example, although $\tau=10$ leads to a higher RMSE difference than $\tau=40$ in most of the cases, $\tau=5$ sometimes even leads to a negative difference value. Under our simulation settings, $\tau=10$ is an acceptable choice.

\subsection{Computation cost}
\label{subsec:compu}
Because there is no privacy-preserving mechanism in TD, we introduce an extended scheme of TD, PPTD~\cite{miao2015cloud} for comparison. PPTD adopts the Threshold Paillier cryptosystem to protect the user privacy.
The computation costs of AirQ and EAirQ are analyzed and compared with that of PPTD in Table~\ref{tab:compu}. 

 \begin{table*}[h]
	\begin{center}
		\renewcommand{\arraystretch}{1.3}
		\centering
		\caption{Computation cost}\label{tab:compu}
		\begin{tabular}{|c|c|c|c|c|}
			\hline
			\multirow{2}{*}{Framework }& \multicolumn{3}{c|}{Vehicles} &\multirow{2}{*}{Cloud server} \\\cline{2-4}
			& Additions &Multiplications & Exponentiations &   \\\hline\hline
			AirQ & $\mathcal{O}(c^2m)$ & $\mathcal{O}(cm)$ & 0 &No additional costs\\ \hline
			EAirQ & $\mathcal{O}(c^2m)$ & $\mathcal{O}(cm)$ & 0 &No additional costs\\ \hline
			PPTD~\cite{miao2015cloud}  & $\mathcal{O}(rm)$  & $\mathcal{O}(rm)$ &$\mathcal{O}(rm)$ & Additional costs  \\
			\hline
		\end{tabular}
	\end{center}
\end{table*}

In AirQ, a vehicle first calculates $\{\theta_{s,j,g}\cdot v_{s,j}\mid  j\in\{1,2,\dots,c\}\}$ and $\{\theta_{s,j,g}\cdot v_{s,j}^2\mid  j\in\{1,2,\dots,c\}\}$ for $g$ as a preparation of the masking process. Thus, there are $2c$ multiplications. Then, $c-1$ additions are needed to mask each value by Equation~\eqref{equ_maskingour01}. For all the $3c$ values, i.e.,  $\{\theta_{s,j,g}\cdot v_{s,j}\mid  j\in\{1,2,\dots,c\}\}$, $\{\theta_{s,j,g}\cdot v_{s,j}^2\mid  j\in\{1,2,\dots,c\}\}$ and $\{\theta_{s,j,g}\mid  j\in\{1,2,\dots,c\}\}$, $3c(c-1)$ additions are needed in the masking process. A strength of AirQ is that vehicles do not need to participate in the following processes after uploading the masked data. Thus, each vehicle only performs $\mathcal{O}(c^2m)$ additions and $\mathcal{O}(cm)$ multiplications in each sensing cycle where $m$ is the number of grids.

In EAirQ, besides the $\mathcal{O}(c^2m)$ additions and $\mathcal{O}(cm)$ multiplications, a vehicle $s$ also needs to add noise to the observation values. As described in Section~\ref{subsec:perturbation}, new observation values are expected to be generated and perturbed for the $m-c$ grids not passed with a probability $p_2$ by Equation~\eqref{equ_rand_respon2} and then all the observation values will be perturbed by Equation~\eqref{equ_noise}. Thus, there are $p_2(m-c)+c$ addition operations for $s$. Overall, a vehicle calculates $\mathcal{O}(c^2m)$ additions and $\mathcal{O}(cm)$ multiplications in EAirQ.
Note that the computation cost of the anonymous authentication is not included in this section because there is no specific signing and verification scheme adopted in PPTD. From the perspective of security, any messages sent should be signed and encrypted so that it is necessary to introduce an authentication scheme in PPTD in practice. Based on the analysis in~\cite{moni2020scalable}, the cost of signing is acceptable in EAirQ.

In PPTD, there are multiple updating rounds for truth discovery until a convergence criterion is satisfied, which is similar to the cases in AirQ and EAirQ. However, the sources should participate in every round to fulfill the truth discovery in PPTD. (Recall that a source is a vehicle in our work but PPTD is not designed particularly for vehicular networks. Thus, we prefer to use the term ``source" instead of ``vehicle" when discussing PPTD.)
In each round, some preparation operations, threshold encryption and decryption are needed. Because the processes are complex, we give the computation cost directly: $\mathcal{O}(mr)$ additions, $\mathcal{O}(mr)$ exponentiations and $\mathcal{O}(mr)$ multiplications for each source where $r$ is number of updating rounds. Note that $m$ is the number of grids but also the number of reports in PPTD because a source is assumed to provide observation values for all grids in PPTD. Some operations are in different fields (such as a multiplicative group) in PPTD but we do not distinguish them in this paper for simplicity.
Readers can refer to~\cite{miao2015cloud} for more details..

Considering the server in our frameworks is configured on the cloud with sufficient calculation resources, we only give a brief discussion on the cost of the server.
As shown in Table~\ref{tab:compu}, except the necessary calculations for truth discovery, there is no additional computation costs in AirQ and EAirQ. In other words, there is no need to remove the randomness from the masked or perturbed values. The server can update the truths and weights based on Equations~\eqref{equ_truth_updating_transfer}, \eqref{equ_weight_updating_transfer}, \eqref{equ_truth_updating2} and \eqref{equ_weight_updating2}
directly. 
However, in PPTD, the server should perform some multiplications and exponentiations to handle the ciphertexts first before using them in the truth and weight updating. 

Overall, AirQ and EAirQ are lightweight, especially on the vehicle-side.

\subsection{Communication cost}
We analyze the communication cost in three aspects as follows.

\textbf{Communication rounds}: in PPTD, there are $\mathcal{O}(r)$ updating rounds for truth discovery in one sensing cycle. In each round, a source should communicate with the server at least twice. The messages sent in one round are $\mathcal{O}(m)$ ciphertexts. Thus, a source sends $\mathcal{O}(rm)$ ciphertexts in each sensing cycle. The size of a ciphertext depends on the parameters chosen for the threshold cryptosystem. 
In both AirQ and EAirQ, there are $c+1$ communication rounds in total in each sensing cycle.
In each of the $c$ rounds, a vehicle requests $\mathcal{O}(m)$ parameters from an RSU as described in Section~\ref{subsub:datagene}.
In the last round, the vehicle uploads the report containing $\mathcal{O}(cm)$ masked or perturbed values. 
Overall, suppose the cost of sending one value (a ciphertext, a parameter, a masked value or a perturbed value) is $\mathcal{O}(1)$. Then the communication cost for a source is $\mathcal{O}(rm)$ in PPTD and $\mathcal{O}(cm)$ in AirQ and EAirQ.
Considering that the communication technologies of vehicular networks are improving fast, it should not be a bottleneck.

\textbf{Communication architectures}: AirQ and EAirQ are more suitable for VANETs from the perspective of communication architectures. To be specific, sources in PPTD need to communicate with the server directly, which results in a long communication time. In AirQ and EAirQ, RSUs are intermediaries to forward the messages between vehicles and the server. Thus,  vehicles do not need to wait for the reply from the server, which is an advantage of AirQ and EAirQ.

\textbf{Traffic bursts}: in PPTD, a source needs to conduct the $\mathcal{O}(r)$ communication rounds in a short period when the server asks for data updating and truth discovery. However, the $c+1$ communication rounds of AirQ and EAirQ are scattered in the whole sensing cycle. Only the last round is for data updating. Thus, it will not lead to bursts and high overhead for the communication network.

\subsection{Privacy analysis}
\label{subsec:privacy}
\subsubsection{Privacy in AirQ}
To protect the privacy, we adopt a masking algorithm in AirQ. The security of the algorithm is threefold: 1) the construct of the mask hides all information of individual inputs except for their sum. 2) The PRNG algorithm provides \emph{pseudo-randomness} to the chosen values (or masks). 3) The masks are used as one-time pads, which provides \emph{true randomness} for the algorithm.

\textbf{Privacy of observation values}: the privacy of observation values is guaranteed by the masking algorithm. In the process of data uploading, each observation value $v_{s,j}$ is not sent directly but with masks, as described in Section~\ref{subsec_datamasking}. In the data handling process, the cloud server can obtain the sums of $\{\theta_{s,j,g}\cdot v_{s,j}\mid  j\in\{1,2,\dots,c\}\}$ and $\{\theta_{s,j,g}\cdot v_{s,j}^2\mid  j\in\{1,2,\dots,c\}\}$ by summing the masked values, $\{\beta_{1}^{s,j,g}\mid j\in\{1,2,\dots,c\}\}$ and $\{\beta_{2}^{s,j,g}\mid j\in\{1,2,\dots,c\}\}$, respectively. In other words, featured by the masking technique, there is no need for the server to acquire the masks (i.e., the chosen random values) or any $v_{s,j}$.

In addition, the masks used in constructing $\{\beta_{1}^{s,j,g}\mid j\in\{1,2,\dots,c\}\}$ and $\{\beta_{2}^{s,j,g}\mid j\in\{1,2,\dots,c\}\}$ are chosen and only known by the vehicle $s$. They work as one-time pads, i.e., they are never reused in other sensing cycles. Thus, it is infeasible for any other parties except $s$ to remove the randomness and infer any value of $V_{s}$.

\textbf{Privacy of vehicle trajectories}: we first emphasize that the server only use Equations~\eqref{equ_truth_updating_transfer} and~\eqref{equ_weight_updating_transfer} for truth discovery. It does not need to know the exact information of the locations of each report in addition.
Then, the only parameter that may disclose the information of locations is $\theta_{s,j,g}$. It represents the logical distance of grids $g^{s,j}$ and $g$. However, all $\{\theta_{s,j,g}\mid  j\in\{1,2,\dots,c\}\}$ are masked to $\{\beta_{2}^{s,j,g}\mid j\in\{1,2,\dots,c\}\}$ and $\{\beta_{3}^{s,j,g}\mid j\in\{1,2,\dots,c\}\}$ for each grid $g$ by source $s$. As mentioned in Section~\ref{subsec:datahandling}, the masked values are sent to the server in reports and used to calculate the sums needed.
It is infeasible for any other parties except $s$ to obtain the exact value of any $\theta_{s,j,g}$ from the reports or the sums. Thus, the privacy of trajectories is preserved.

As discussed in our preliminary work~\cite{liu2020airq}, 
although AirQ avoids the disclosure of trajectories from the reports sent to RSUs and the cloud server, the periodic communications between vehicles and RSUs bring risks to the privacy. This remaining challenge is tackled in this work with EAirQ. Details are given in Section~\ref{subsec:priv_EAirQ}.

\subsubsection{Privacy in EAirQ}
\label{subsec:priv_EAirQ}
To protect the privacy of vehicles, we present a two-layer perturbation mechanism and adopt an anonymous communication scheme in EAirQ. We first describe how the perturbation mechanism protects the privacy:

\textbf{Privacy of observation values}: in the value perturbation process, every observation value is perturbed first with a Laplace noise by the vehicle $s$ locally before uploading. It is infeasible to guess the exact value of the noise.
Thus, the true observation values are never disclosed to any parties except $s$. 

\textbf{Privacy of vehicle trajectories}: in the grid perturbation process, every grid that a vehicle $s$ passes by is expected to be concealed with $p_1$. It is infeasible for any parties except $s$ to know which trajectory is removed.
Besides, $(m-c)p_2$ imitated observation values are expected to be generated, where $c$ is the number of observation values provided before the perturbation. Thus, each location inferred from the report has a probability of $\frac{(m-c)p_2}{(m-c)p_2+c}$ to be false. Recall that a Laplace noise is added to the imitated observation values by Equation~\eqref{equ_rand_respon2}, which makes it challenging to judge if a value is manually generated from the historical truth. In other words, it is challenging to distinguish a false trajectory from the true trajectory.

As for the anonymous communication, 
the privacy properties of it are as follows: 1) a vehicle $s$ never uses its real ID $RID_s$ to upload reports. Both the RSUs and the server cannot infer the true identity of $s$. 2)  $s$ can use any of the many existing pseudonyms changing techniques to change its pseudonym between the sensing cycles~\cite{moni2020scalable}. 3) Only the TM knows the $RID_s$. When the server requires the historical weights for $s$ with the pseudo-ID $PID_s$ from the TM, only the weights are returned. 4) The weight histories maintained by the TM are expected to be changing based on the application-layer user activities, which increases the challenges to link a record with a vehicle by the server. The four properties guarantee that it is infeasible for the server to infer the relationship between a report with a vehicle. 
It also overcomes the remaining privacy limitation in AirQ. In other words, the RSUs cannot trace a vehicle in the periodic communications.
Thus, both the observation values and the trajectories are protected.

Note that AirQ and EAirQ preserve the privacy of observation values and vehicle trajectories while PPTD preserves that of observation values and weights. The difference in privacy goals is reasonable considering the practical scenarios and applications. 

%% file: EAirQ_6_Dis.tex
\section{Discussion}
 \label{sec:dis}

 \textbf{Possible scenarios}: 
 vehicles with high mobility can collect various data from the environment, such as the noise level, the humidity, the temperature, and the flow density. 
 Thus, AirQ and EAirQ are not restricted to air quality monitoring. They can also adapt to other scenarios well. 
 The remaining work is to adjust the parameters based on the degrees of temporal and spatial correlations.  
 For example, we can decrease $\rho_t$ in Equation~\eqref{equ_truthcorr} properly for the sensing task with a significant temporal correlation on ground truths such as the outdoor temperature. Similarly, the threshold $u$ and width parameter $\omega$ in Equation~\eqref{equ_gaussian} should be adjusted based on the spatial property of the sensing object. To emphasize that, although the parameters are expected to be changed for different applications, we do not need to modify the formulas for truth discovery. 
 
 \textbf{A possible extension of correlations}: 
 in this paper, only the temporal and spatial correlations are involved, but we discuss the possibility of the \emph{attributes-based correlation}. Intuitively, 
 two grids are likely to have similar air quality values if they are similar in some particular attributes. For instance, the construction sites with heavy-duty engines are more likely to produce diesel emissions.
 We define the similarity of $g^{s,t}$ and $g$ as:
  \begin{equation}
 \label{equ_LWD}
 \Sim(g^{s,j}, g) =
 \left\{
 \begin{array}{lr}
 \begin{split}
 1-\Nor(\LWD\\(L_{g^{s,j}}, L_g)), 
 \end{split}
 &  \begin{split} \text{if}\; \Nor(\LWD(L_{g^{s,j}},\\ L_g)) < u^\prime  \end{split}\\
 0, & \text{otherwise}
 \end{array} \right.
 \end{equation}
 where $L_g=\{f_1,f_2,\dots,f_z\}$ is the vector of $z$ attributes of grid $g$, including the vehicle density, the vegetation ratio, the number of factories, the duration of constructions and so on.
 Similarly, $L_{g^{s,j}}=\{f_1^\prime,f_2^\prime,\dots,f_z^\prime\}$ is the attribute vector of $g^{s,t}$.
  $\LWD(L_{g^{s,j}},L_g)$ is the \emph{Lance and Williams distance} of two vectors $L_{g^{s,j}}$ and $L_g$. The Lance and Williams distance is also called \emph{Canberra distance}, which assumes that the variables in a vector are independent of each other. It is widely used to measure the similarity and the dissimilarity between groups.
  To be specific, $\LWD(L_{g^{s,j}},L_g)$ can be calculated as follows,
  \begin{equation}
 \LWD(L_{g^{s,j}},L_g) = \sum_{i=1}^z \frac{\left| f_i^\prime - f_i \right |}{\left| {f^{\prime}_i}\right| + \left| f_i \right|}
 \end{equation}
 
  Recall that the attributes of all the grids can be known in advance by the RSUs. Thus, the distances between the attribute vectors can be calculated and normalized by the RSUs in advance as well. Note that the normalization is processed among all the distances but we use $\Nor(\LWD(L_{g^{s,j}},L_g))$ in Equation~\eqref{equ_LWD} to denote the normalization result of $\LWD(L_{g^{s,j}},L_g)$.
  $\Nor(\LWD(L_{g^{s,j}},L_g))\in\left[0, 1\right]$. 
  If two grids are similar, they are expected to have a small Lance and Williams distance and the normalization result tends to 0. Thus, the similarity is finally defined as 
 $1-\Nor(\LWD(L_{g^{s,j}},L_g))$ when a distance threshold $u^\prime$ is satisfied. 
 
 
 The parameter $\theta_{s,j,g}$ then can be redefined as 
 \begin{equation}
 \theta_{s,j,g} = \psi_1 \Dis(g^{s,j}, g) +  \psi_2 \Sim(g^{s,j}, g) 
 \end{equation}
 where $\psi_1$ and $\psi_2$ control the weights of the two correlations. 
 The challenge of adopting the attributes-based correlation is the complexity of attributes.
 
 \textbf{The trade-off between accuracy and privacy in EAirQ}: we have mentioned that adding more noise to reports can provide a higher privacy but may lead to a lower accuracy. A system manager can weigh it according to specific demands. For example, a system that needs results of concrete values should have less noise added than a system that only needs classified outputs (such as \emph{Heavily polluted}). 
 
 The parameters $p_1$, $p_2$, $\lambda_1$ and $\lambda_2$ can be adjusted accordingly. As observed from Fig.~\ref{fig:per_EAirQ_para_c}, $\lambda_2$ can be set larger than what we use in most of the simulations (i.e., 2). It provides more privacy but does not affect the accuracy much, which is the strength of the Laplace noise. However, one point need to be discussed is the weight issues introduced by it. With more noise added to the observation values, the estimated weights for sources are deviated more from the real reliabilities. If the estimated weights are only used for truth discovery, it is acceptable to use a larger $\lambda_2$. If the system manager uses the weights for some application-layer functions as well, such as giving rewards to sources based on their weights, we suggest keeping $\lambda_2$ small.
 
  \textbf{Weight management in EAirQ}: recall that the TM has the ability to update the historical records of weights so that the changing records provide better protection of the privacy. However, because it is not the main focus of our work, we do not discuss much and there are many remaining problems. For example, how to define a bad source with the weight? In other words, what should be the threshold of the weight for a bad source? How to reasonably update the whole $W_s$ for each $s$ rather than only the latest $w_{s,t-1}$ by the TM? A concrete list of updating and management rules should be proposed in the future.
  
  \textbf{Datasets and simulations}: the simulations conducted in this work are not perfect. 
  We adopt a dataset and manually generate some data such as the observation values. The dataset is the most fine-grained one that we can find. However, it is not enough. Getting a dataset with the truths of streets or blocks is challenging in practice. We believe the limitation of datasets affects the performance shown in the simulations. Besides, if it is possible to conduct experiments with sufficient volunteers in practice, a more dependable evaluation can be provided. Based on our observations, the discussed challenges are common in similar crowdsensing research. 
  
  \textbf{Vehicle mobility}: recall that we set up simplified trajectories for vehicles, instead of generating real trajectories with concrete mobility settings. 
  The reasons that we do not adopt a concrete mobility model in this work are twofold: 1) without considering the communication performance (such as the packet loss rate), the vehicle mobility does not affect the truth discovery performance. Vehicles only upload the reports to the nearest RSU in each sensing cycle. After that, the vehicles do not need to participate in any following data handling process. Thus, they can travel to any destination at any speed; 2) definitely the vehicle movements have an influence on the communication performance in VANETs. However, the corresponding QoS is not the main focus of this paper, as assumed in Section~\ref{sebsec:prb_def}.
  As a future work, some complex mobility models~(such as that in~\cite{qian2017dynamic}) and the effects on the QoS can be taken into consideration. Besides, whether and how the mobility, especially the speed, affects the precision of the onboard sensors is also a potential research topic.

 \section{Conclusion}
 \label{sec:con}
 In this paper, we presented a truth discovery algorithm for vehicular crowdsensing incorporating the spatial and temporal correlations. We further proposed a lightweight privacy-preserving framework based on data masking. 
 The proposed framework, AirQ, can address the data sparsity problem and preserve the privacy of reports and trajectories at the same time. 
 Thus, it is suitable for fine-grained tasks such as air quality monitoring. 
 In addition, an enhanced version of AirQ, EAirQ was presented with the techniques of anonymous communication and perturbation. The goals of EAirQ are threefold: 1) to overcome the limitations of AirQ in truth discovery; 2) to protect the privacy of reports and trajectories; 3) to be lightweight for VANETs.
 We conducted simulations to measure the performance and analyzed the privacy achievements of the two frameworks. 
 Results show that both EAirQ and AirQ protect the privacy while keeping the computation and communication costs low, and EAirQ performs better than AirQ in truth discovery accuracy.